\documentclass[a4paper,11pt]{article}
\pdfoutput=1 

\usepackage{jheppub} 

\def\sfrac#1#2{{\textstyle{#1\over #2}}}
\newcommand{\be}{\begin{equation}}
\newcommand{\ee}{\end{equation}}
\newcommand{\ba}{\begin{array}}
\newcommand{\ea}{\end{array}}
\newcommand{\bea}{\begin{eqnarray}}
\newcommand{\eea}{\end{eqnarray}}
\newcommand{\sss}{\scriptscriptstyle}

\newcommand{\nn}{\nonumber}

\newcommand{\W}{{\sss W}}

\title{\boldmath A little theory of everything, with heavy neutral leptons}

\author[a,1]{James Cline,\note{Corresponding author.}}
\author[a]{Matteo Puel}
\affiliation[a]{McGill University, Department of Physics, 3600 University St.,
Montr\'eal, QC H3A2T8 Canada}
\author[b]{and Takashi Toma}
\affiliation[b]{Institute of Liberal Arts and Science, Kanazawa University, Kakuma-machi, Kanazawa, Ishikawa 920-1192 Japan}

\emailAdd{jcline@physics.mcgill.ca}
\emailAdd{matteo.puel@mail.mcgill.ca}
\emailAdd{toma@staff.kanazawa-u.ac.jp}

\abstract{Recently a new model of ``Affleck-Dine inflation'' was
presented,  that produces the baryon asymmetry from a complex inflaton
carrying baryon  number, while being consistent with constraints from
the cosmic microwave background. We adapt this model such that the
inflaton carries lepton number, and communicates the lepton asymmetry
to the standard model baryons  via quasi-Dirac heavy neutral leptons
(HNLs) and sphalerons. One of these HNLs, with mass $\lesssim
4.5\,$GeV, can be (partially) asymmetric dark matter (DM), whose
asymmetry is determined by that of the baryons. Its stability is
directly related to the vanishing of the lightest neutrino mass.  
Neutrino masses are generated by integrating out heavy sterile
neutrinos whose mass is above the inflation scale. The model provides
an economical origin for all of the major ingredients missing from the
standard model: inflation, baryogenesis, neutrino masses, and dark
matter. The HNLs can be probed in fixed-target experiments like  SHiP,
possibly manifesting $N$-$\bar N$ oscillations. A light singlet
scalar, needed for depleting the DM symmetric component, can be
discovered in beam dump experiments and searches for rare decays,
possibly explaining anomalous events recently observed by the KOTO
collaboration. The DM HNL is strongly constrained by direct searches,
and could have a cosmologically interesting self-interaction cross
section.}

\begin{document} 
\maketitle
\flushbottom

\section{Introduction}

The standard model (SM) of particle physics is noted for being incomplete
in numerous ways.  It could be argued that the most urgently missing 
elements are an inflaton (or other source of primordial density
perturbations), a mechanism for baryogenesis, dark matter (DM), and the
origin of neutrino masses, since all of these relate to directly 
observed phenomena as opposed to problems of naturalness.  It is
tempting to seek relatively simple new physics models that can
simultaneously address several of the missing pieces, or perhaps all.\footnote{Ref.\ \cite{Alonso-Alvarez:2019fym} provides a recent attempt in this direction, in which an inflaton-like field is present, although the details of inflation are not yet worked out.}

A notable example is the $\nu$MSM 
\cite{Asaka:2005pn,Shaposhnikov:2006xi}, in which light sterile
neutrinos can accomplish leptogenesis and provide a dark matter
candidate while giving neutrino masses.  Higgs inflation 
\cite{Bezrukov:2007ep} can be invoked in this framework without needing any additional particles.  
A similar mechanism of getting an inflationary phase was also implemented in the scotogenic model \citep{Choubey:2017hsq,Borah:2018rca} to simultaneously explain inflation, dark matter, baryogenesis and neutrino masses, by introducing a scalar inert doublet coupled non-minimally to gravity and three right-handed neutrinos.
Another example is the SMASH model \cite{Ballesteros:2016euj} that assumes heavy right-handed neutrinos to explain neutrino mass and thermal leptogenesis, while introducing minimal extra matter content to produce axions as dark matter and a solution to the strong CP problem.  The extra scalar field needed for breaking Peccei-Quinn symmetry can combine with the Higgs to give two-field inflation in the early universe. 
The idea of explaining neutrino masses, baryon asymmetry, dark matter, inflation and solving the strong CP problem using three right-handed neutrinos and the extra fields of the KSVZ axion model \cite{Shifman:1979if} was originally presented in ref.\ \cite{Salvio:2015cja}. There, the Higgs field was identified as the inflaton and the electroweak vacuum was shown to be stable for several choices of the model parameters. The problem of Higgs inflation, which is known to reduce the scale of perturbative unitarity breaking well below the Planck scale, was addressed by coupling the Higgs field nonminimally to gravity \cite{Salvio:2018rv}.  

In the present work we suggest another way of completing the standard
model, that does not rely upon leptogenesis as usually defined
(through the CP-violating out-of-equilibrium decays of heavy
neutrinos).  The starting point is a model of inflation in which the
Affleck-Dine mechanism \cite{Affleck:1984fy} for creating a particle asymmetry occurs
{\it during} inflation \cite{Cline:2019fxx}.  The asymmetry is
originally stored in a complex inflaton field, that has the Lagrangian
\be
	{\cal L} = {m_P^2\over 2}R \left(1 +
	2\xi|\phi|^2\right) + 
	|\partial\phi|^2 - m_\phi^2|\phi|^2 - \lambda|\phi|^4 -
	i\lambda'(\phi^4-\phi^{*4})
\label{lagr}
\ee
(where $m_P$ is the reduced
Planck scale) including a nonminimal coupling to gravity , needed to flatten the
potential at large $|\phi|$, which makes the inflationary predictions
compatible with Planck constraints \cite{Akrami:2018odb}.  In ref.\ 
\cite{Cline:2019fxx} we assumed that $\phi$ carried baryon number,
which was transferred to the SM quarks through colored scalar
mediators.  Here we consider the case where $\phi$ carries lepton
number, hence giving a new mechanism of leptogenesis.  As usual, the
resulting lepton asymmetry is transmitted to the baryons through the
sphaleron interactions of the SM.

The challenge for such an approach is to find a way of transferring
the lepton asymmetry from $\phi$ to the SM without it being washed out
by the lepton-violating effects associated with neutrino mass
generation.  Indeed, if $\phi$ decays to heavy right-handed neutrinos
that have large Majorana masses, the asymmetry gets washed out
immediately and the situation reverts to standard leptogenesis being
required.  This suggests that $\phi$ should decay into quasi-Dirac
neutrino mediators $N_i$, that mix with the SM neutrinos to transmit
the asymmetry.  Among the $N_i$ mediators, one can be stable and
constitute a species of asymmetric dark matter, getting its relic
density (partly) from the initial lepton asymmetry.  The $N_i$ are an example
of heavy neutral leptons (HNLs), a class of hypothetical particles 
that is being widely studied both theoretically and by upcoming
experiments such as SHiP \cite{Alekhin:2015byh}, 
MATHUSLA \cite{Curtin:2018mvb}, FASER \cite{Feng:2017uoz} and CODEX-b \cite{Gligorov:2017nwh}.

To deplete the symmetric component of the DM to a viable level, it is necessary to
introduce a light mediator, which we take to be a scalar singlet $s$,
so that $N_i\bar N_i\to ss$ annihilations are sufficiently strong. 
The DM can be fully or partially asymmetric depending on the coupling
strength $g_s$.  We will show that this interaction has interesting
implications for direct detection, and for hints of anomalous rare
$K_L\to\pi^0$+ invisible decays that have recently been reported by the
KOTO experimental collaboration \cite{Shinohara}.

In our proposal, the HNLs do not explain the origin of light neutrino
masses, but we hypothesize that their couplings to the SM $\nu$'s are
related to those of the superheavy Majorana $\nu_R$'s that generate
seesaw masses, by a principle similar to minimal flavor violation
(MFV) \cite{DAmbrosio:2002vsn}.  The setup thereby also addresses
the origin of neutrino mass and relates the HNL couplings to it in
an essential way.  Moreover a direct link is made between the
stability of the dark matter candidate and the masslessness of the
lightest SM neutrino.

In section \ref{model} we specify the structure of couplings of the HNLs to the 
inflaton and SM particles, and its relation to neutrino mass
generation.  In section\ \ref{leptogen} we discuss constraints on the
couplings such that the lepton asymmetry from inflation is transferred
to the SM particles without being washed out.  It is shown how the
resulting baryon asymmetry determines the dark matter asymmetry and 
its mass.  The relations between light $\nu$ properties and the HNL
couplings are presented in section\ \ref{neutrino}, and consequent
predictions for the phenomenology of the HNLs.  In section\
\ref{singlet-const} we compile the experimental limits on the light
singlet $s$, and identify a region of parameter space where the KOTO
anomaly can be reconciled with DM direct detection limits.  The
latter are considered in detail in section\ \ref{ddsect}, where we also treat the DM self-interactions and discuss possible DM indirect detection constraints.  The technical naturalness of our 
setup is demonstrated in section\ \ref{naturalness}, followed by 
conclusions in section\ \ref{conclusions}.  In appendix \ref{appA}
we derive the exact width for HNL decay into different-flavor
charged leptons, which was given only in approximate form in previous
papers.

\section{Model}
\label{model}
We assume the inflaton carries lepton number 2 (more correctly,
$B-L = -2$ since $B-L$ symmetry is not broken by electroweak
sphalerons), and couples to 
$N_N$ flavors of quasi-Dirac HNLs as 
\be
	g_\phi \phi \bar N_{L,i} N_{L,i}^c + 
	g_\phi \phi \bar N_{R,i} N_{R,i}^c
	+ {\rm H.c.}
\label{Nphi}
\ee
$N_N$ is a free parameter; hereafter we take
$N_N = 3$, which is the minimal number
needed to get dark matter and the observed neutrino properties,
through consistent assumptions about the flavor structure of the
neutrino sector that will be explained presently.
The HNLs couple to the SM lepton doublets as
\be
	\eta_{\nu,ij} \bar N_{R,i} H L_j 
\label{etaeq1}
\ee

At energy scales relevant for inflation and below, it is consistent to
assume that the only source of lepton number violation is
through a small Majorana mass $\epsilon_\nu$ for the standard model
neutrinos, which could be generated through the seesaw mechanism, by integrating
out very heavy right-handed neutrinos, with mass $M_{\nu_R}$ above the scale of
inflation.
In the basis $\nu_L, N_R^c, N_L$, the neutrino mass matrix is
\be
	\left(\begin{array}{ccc} \epsilon_\nu & \eta_\nu^T\, \bar v & 0  \\
				\eta_\nu\, \bar v & 0 &  M_N  \\
	                        0 & M_N & 0 \\
	\end{array}\right)
\label{33mm}
\ee
where $\bar v\cong 174$\,GeV is the complex Higgs VEV.  
We assume that $\epsilon_\nu$ has
a flavor structure that is aligned with the couplings in (\ref{etaeq1})
as
\be
	\epsilon_\nu = \bar\mu_\nu\, \eta_\nu^T\eta_\nu
\label{epsrel}
\ee
where $\bar\mu_\nu$ is a scale that we will constrain below.  This alignment 
ensures the stability of dark matter against oscillations 
with its antiparticle, if $\eta_\nu$ has one
vanishing eigenvalue.  In order to justify the ansatz, we will show
that it is radiatively stable, due to an
approximate SU(3) flavor symmetry for the $N_i$ leptons, that is
broken in a minimal-flavor-violating (MFV) \cite{DAmbrosio:2002vsn} 
manner, solely by the 
matrix $\eta_\nu$.  For example, the flavor-diagonal
couplings of the inflaton to
$N_i$ could be perturbed by a term proportional to $\eta_\nu\eta_\nu^T$ without spoiling
the viability of the framework. 

By solving for the eigenvalues of 
(\ref{33mm}), one finds that the light neutrino part $\epsilon_\nu$ induces a small Majorana mass matrix
for the $N_i$'s of the form
\be
	\delta M = {\bar v^2\over M_N^2}\, \eta_\nu\, 
\epsilon_\nu\, \eta_\nu^T
\label{dMeq}
\ee
that leads to $N_i$-$\bar N_i$ oscillations.  These are mildly constrained
by the need for approximate lepton number conservation during the
generation of the lepton asymmetry (apart from 
electroweak sphalerons), as we consider below.  

\section{Nonstandard leptogenesis and DM relic density}
\label{leptogen}

During inflation $\phi$ gets an asymmetry determined mostly by the couplings
in eq.\ (\ref{lagr}) and to a smaller extent by the initial conditions of the inflaton,
which provide the source of CP violation in the Affleck-Dine mechanism
\cite{Affleck:1984fy}.  The details of asymmetry generation
at the level of $\phi$ are
exactly the same as discussed in ref.\ \cite{Cline:2019fxx}.
The difference in the present work is that the $\phi$ asymmetry is transferred to the HNLs by the decays $\phi\to NN$ from the
interaction (\ref{Nphi}).  
Whether reheating is perturbative or 
proceeds by parametric resonance is not crucial to the present
discussion, where we assume that the created asymmetry results in the
observed baryon asymmetry.  This can always be achieved by appropriate
choice of the $L$-violating parameter $\lambda'$, for example.\footnote{This observation is consistent with the results obtained by including the effects from nonlinear preheating dynamics on the generation
of matter-antimatter asymmetry in Affleck-Dine
inflationary scenarios\ \cite{Lozanov:2014zfa}.}

\subsection{Sharing and preserving the asymmetry}
\label{sharing}

For simplicity, consider the case where 
$g_\phi$ is sufficiently small so that perturbative
decays are the dominant mechanism for reheating, with 
reheat temperature of order
\be
	T_R \sim g_\phi (m_\phi m_P)^{1/2}\sim 10^{-3} g_\phi\, m_P
\ee
using the typical value $m_\phi\sim 10^{-6}\,m_P$ identified in
ref.\ \cite{Cline:2019fxx}. 
Even for rather small values $g_\phi \lesssim 0.01$, this is well 
above the weak scale.  Therefore it is easy for the HNLs to 
equilibrate with the SM through the interaction (\ref{etaeq1}), which
transmits the primordial $B-L$ asymmetry to the SM. The dominant
process is $N_i$ (inverse) decays, whose rate is 
$\Gamma_{d}\cong 10^{-3}\eta_\nu^2 T$ \cite{Campbell:1992jd}
for $T\gtrsim 100$\,GeV.  Demanding that this comes into equilibrium
before sphalerons freeze out, we find the lower bound
$|\eta_\nu| \gtrsim 4\times 10^{-7}$ on the largest elements of 
$\eta_{\nu,ij}$.

We demand that no $L$-violating effects from the operator $\lambda'\phi^4$
in eq.\ (\ref{lagr}) ever come into equilibrium, since these would wash out
the asymmetry.  Above the scale $m_\phi$, this comes from $\phi\phi\to 
\phi^*\phi^*$ scatterings with rate $\sim \lambda'^2\, T$, that comes into 
equilibrium at $T\sim \lambda'^2 m_P\sim 10^{-24}m_P$, using the typical
value $\lambda'\sim 10^{-12}$ found in ref.\ \cite{Cline:2019fxx}.  This
is far below $m_\phi$, hence it never comes into equilibrium.  Instead the principal
effect of $\lambda'$ is through the effective operator $(\lambda'g_\phi^4/m_\phi^8)
(\bar N N^c)^4$ generated by integrating out the inflaton.  This has
a rate going as $\lambda'^2 g_\phi^8 m_\phi^{-16}T^{17}$, that goes out of equilibrium
at $T\sim [m_\phi^{16}/(\lambda'^2 g_\phi^8 m_p)]^{1/15}$.  Demanding that this
remains below the reheat temperature gives an upper bound on $g_\phi$,
\be
	g_\phi \lesssim \left(m_\phi\over m_P\right)^{17/23}
\left(1\over \lambda'\right)^{2/23} \cong  0.07
\ee
which is not prohibitive.

The only other $L$-violating process operative at scales below that of
inflation is $N$-$\bar N$ oscillations induced by the 
$\delta M$ matrix elements (\ref{dMeq}).  These would 
wash out the $B$ and $L$ asymmetries if they were in equilibrium
before sphaleron freezeout.  The rate of $L$ violation is not 
simply the same
as the oscillation rate $\sim 1/\delta M$, because flavor-nondiagonal
interactions of $N$ with the plasma can measure the state of the
oscillating $N$-$\bar N$ system before it has time to oscillate 
significantly, damping the conversions of $N\to\bar N$.  The
effective rate of $L$ violation can be parametrized as
\cite{Cline:1991zb,Bringmann:2018sbs}
\be
	\Gamma_{\Delta L} \sim {M_N^2\delta M^2\over M_N^2\delta M^2
	+ T^2\Gamma_{m}^2}\,\Gamma_{m}
\label{GDL}
\ee
where $\Gamma_m$ is the rate of processes that destroy the
coherence of the $N$-$\bar N$ system.\footnote{We will introduce an additional elastic scattering channel mediated
by a singlet scalar $s$ below.  These flavor-conserving interactions are not relevant
for decohering the $N$-$\bar N$  oscillations \cite{Tulin:2012re}.}\ \
  For $T> T_{\rm EW}\sim
100$\,GeV, (inverse) decays are dominant, but these quickly go out
of equilibrium as $T$ falls below the mass of the Higgs boson.
At temperatures somewhat below $T_{\rm EW}$, the elastic (but
flavor-violating)  $NL\to \bar NL$ scatterings mediated by Higgs exchange
dominate, with $\Gamma_m = \Gamma_{\rm el} \sim \eta_\nu^4 T^5/m_h^4$.  On the
other hand, sphalerons are safely out of equilibrium since they are
exponentially suppressed by the Boltzmann factor involving the
sphaleron energy, which is above the TeV scale.  
Therefore it is sufficient to show that the rate (\ref{GDL}) is out of equilibrium
in this case, to establish that the washout process is innocuous.
%
%
In other words, the following relation must be satisfied
\be
\frac{\Gamma_{\Delta L}}{H} \sim \frac{M_N^2 \delta M^2\ m_h^4\ m_P}{\eta_{\nu}^{4}\ T_{\text{EW}}^9} < 1
\label{GammaLoH}
\ee

In section \ref{neutrino} we will show that the light neutrino mass
matrix $m_{\nu}$ is approximately equal to $\epsilon_{\nu}$, which is
generated by integrating out heavy neutrinos though the usual seesaw
mechanism. This allows us to rewrite the HNL Majorana mass matrix
$\delta M$ in eq.\ \eqref{dMeq} as $\delta M\sim \bar{v}^2
\eta_{\nu}^2\ m_{\nu} / M_{N}^2 \sim U_{\ell i}^2\, m_\nu$
where $U_{\ell i}$ is the mixing angle between HNLs and light
neutrinos. Plugging the latter in eq.\
\eqref{GammaLoH}, the $\eta_{\nu}$-dependence disappears and we can
get a lower bound on the HNL Dirac mass, $M_N \gtrsim 4$ MeV. For
higher values of $M_{N}$, the lepton-violating effects of $\delta M$
are therefore too small to affect the baryon asymmetry, but they can
be observable in collider experiments that we will discuss in section
\ref{neutrino}. %

DM-antiDM oscillations for asymmetric DM have been considered in 
refs.\ \cite{Tulin:2012re,Cirelli:2011ac}.  They can potentially
regenerate the symmetric component of the DM and lead to its
dilution through annihilations.
We avoid these
constraints by the relation (\ref{dMeq}) that causes $\delta M$ to vanish 
when acting on  the $N'$ DM state.

\subsection{DM asymmetric abundance and maximum mass}
The relic density for fully asymmetric DM is determined by its
chemical potential, which in our framework is related to the
baryon asymmetry in a deterministic way, since the DM
initially has the same asymmetry as the remaining two HNLs.
The 
relation between the DM and baryon asymmetries can be found by
solving the system of equilibrium constraints, similarly to
ref.\ \cite{Harvey:1990qw}.  We generalize their network to include 
the extra HNL species, that satisfy the equilibrium
condition
\be
	\mu_N = \mu_h + \mu_L
\label{chemeq}
\ee
from the $\eta_\nu$ interactions. Eq.\ (\ref{chemeq}) only
applies to the unstable HNL species since $N'$ is conserved,
and its chemical potential is fixed by the initial lepton asymmetry
\be
	\qquad\qquad\qquad\qquad\quad\mu_{N'} = \sfrac1{6} L_0
	\qquad\qquad\qquad\qquad
\ee
The factor of 6 comes from having three HNL species, each with two
chiralities.  We recall that $L_0$ is determined by the inflationary
dynamics, and is especially sensitive to the value of the coupling $\lambda'$.
It is assumed that $\lambda'$ has been adjusted so that $L_0$ takes the
value needed to yield the observed baryon asymmetry, which we relate
to $L_0$ in the following.

Repeating the analysis of \cite{Harvey:1990qw} we find the following
equilibrium relations (setting the $W$ boson potential $\mu_W =0$
since $T>T_{\text{EW}}$):
\bea
	L &=& \sfrac{13}{3}\mu_\nu + \mu_h +
2\mu_{N'}\nn\\ &=& 
	\sfrac{95}{21}\mu_\nu + 2\mu_{N'}\nn\\
	B &=& -\sfrac43\mu_\nu\nn\\
	\mu_{u_L} &=& -\sfrac19\mu_\nu,\quad
	\mu_h = \sfrac{4}{21}\mu_\nu\nn\\
	\mu_N &=& \sfrac{11}{21}\mu_\nu
\label{chemnet}
\eea
where $L,B$ are the respective total chemical potentials for lepton and baryon
number, $\mu_\nu$ is the sum of light neutrino chemical potentials,
and $\mu_h$ is that of the Higgs.
Since $B-L$ is conserved by sphalerons, we can relate these to the
initial lepton asymmetry $L_0 = 6\mu_{N'}=(L-B)$: 
$\mu_\nu = \sfrac{84}{123}\mu_{N'}$, 
$B=-\sfrac{112}{123}\mu_{N'}$.
This allows us to determine the maximum mass of $N'$ that gives the observed
relic density:
\be
	m_{N'} = M_N \le \left|B\over \mu_{N'}\right|{\Omega_c\over \Omega_b}
	\, m_n 	= 	4.5\,{\rm GeV}
\label{massbound}
\ee
using the values $\Omega_c = 0.265$ and $\Omega_b = 0.0493$ 
from ref.\ \cite{Aghanim:2018eyx} and the nucleon mass $m_n$.

The inequality (\ref{massbound}) is only saturated if the symmetric
DM component is suppressed to a negligible level.  Otherwise a
smaller value of $m_{N'}$ is needed to compensate the presence of
the symmetric component.  We turn to the general case next.

\subsection{Dark matter annihilation and relic density}
\label{dma_rel}

In order to reduce the symmetric component of the DM to avoid
overclosure of the universe, an additional annihilation channel is
needed.
The $t$-channel Higgs-mediated annihilations $N'\bar N'\to L\bar
L$ are not strong enough, leading to $\langle\sigma v\rangle\lesssim 10^{-32}$ cm$^3$/s, 
in light of the bound $|\eta_\nu|\lesssim  10^{-3}$ to be derived in section \ref{neutrino} below.
We need an additional particle with sufficiently strong couplings
to the DM.

The simplest possibility is to introduce a singlet scalar
$s$, with interactions
\be
	g_s s\bar N_i N_i + \sfrac14\lambda_s (s^2 - v_s^2)^2
	+ \lambda_{hs}h^2 s^2
	\label{eq:s_int}
\ee
that at tree level are diagonal in the $N_i$ flavors, and lead to
mixing of $s$ with the Higgs $h$.   We will consider two cases:
(i) $m_s < m_{N'}$ so that $N'\bar N'\to ss$ is allowed; 
(ii) $m_s\gtrsim 2\, m_{N'}$ so that there can be mild resonant
enhancement of the $s$-channel cross section for $N'\bar N'$
annihilation to standard model particles, through the mixing of 
$s$ with the Higgs boson.
For the nonresonant case, the $s$-channel amplitude for $N'\bar N'\to f\bar f$, 
where $f$ is the most strongly coupled
kinematically accessible final state, is of the same order of magnitude
as that for $N'$-nucleon scattering, which is strongly constrained by
direct detection (section \ref{ddsect}), making this contribution 
too small to be sufficient for annihilation.  We will see that
this limitation can be overcome by resonant enhancement without requiring too much fine tuning of masses.
  
\subsubsection{$N'\bar N'\to ss$ annihilation}

We first consider the case when $m_s < m_{N'}$.
The cross section for $N'\bar N'\to ss$ is $p$-wave suppressed.  
Parameterizing the Mandelstam
variable as $s = 4 m_{N'}^2(1 +
\epsilon)$ we find in the limit $m_s\ll m_{N'}$ and $\lambda_s \ll
g_s$ that
\be
	\sigma \cong {3\, g_s^4\over 64\pi\, m_{N'}^2} {\epsilon^{1/2}
\over (1+\epsilon)^2 }
\ee
(this is an analytic approximation to the exact result, which is
more complicated).  Carrying out the thermal average 
\cite{Gondolo:1990dk}
with $x \equiv m_{N'}/T$ gives
\bea
	\langle\sigma v\rangle &\cong& {3\, g_s^4\over 16\pi\, m_{N'}^2}
	F(x)\\
	F(x) &=& {x\over K_2(x)^2}\int_0^\infty d\epsilon 
	\left({\epsilon\over1+\epsilon}\right)^{3/2}\,
	K_1(2x\sqrt{1+\epsilon})\\
	&\cong&0.058 - 0.002\, x + 3.25\times 10^{-5} x^2
 - 1.87\times 10^{-7} x^3\nn
\eea
which is a good numerical approximation in the region $15<x<70$.
For values $x\sim 20$ typical for freezeout, $F \cong 0.03$.

\begin{figure}[t]
\begin{center}
\centerline{
\includegraphics[scale=0.261]{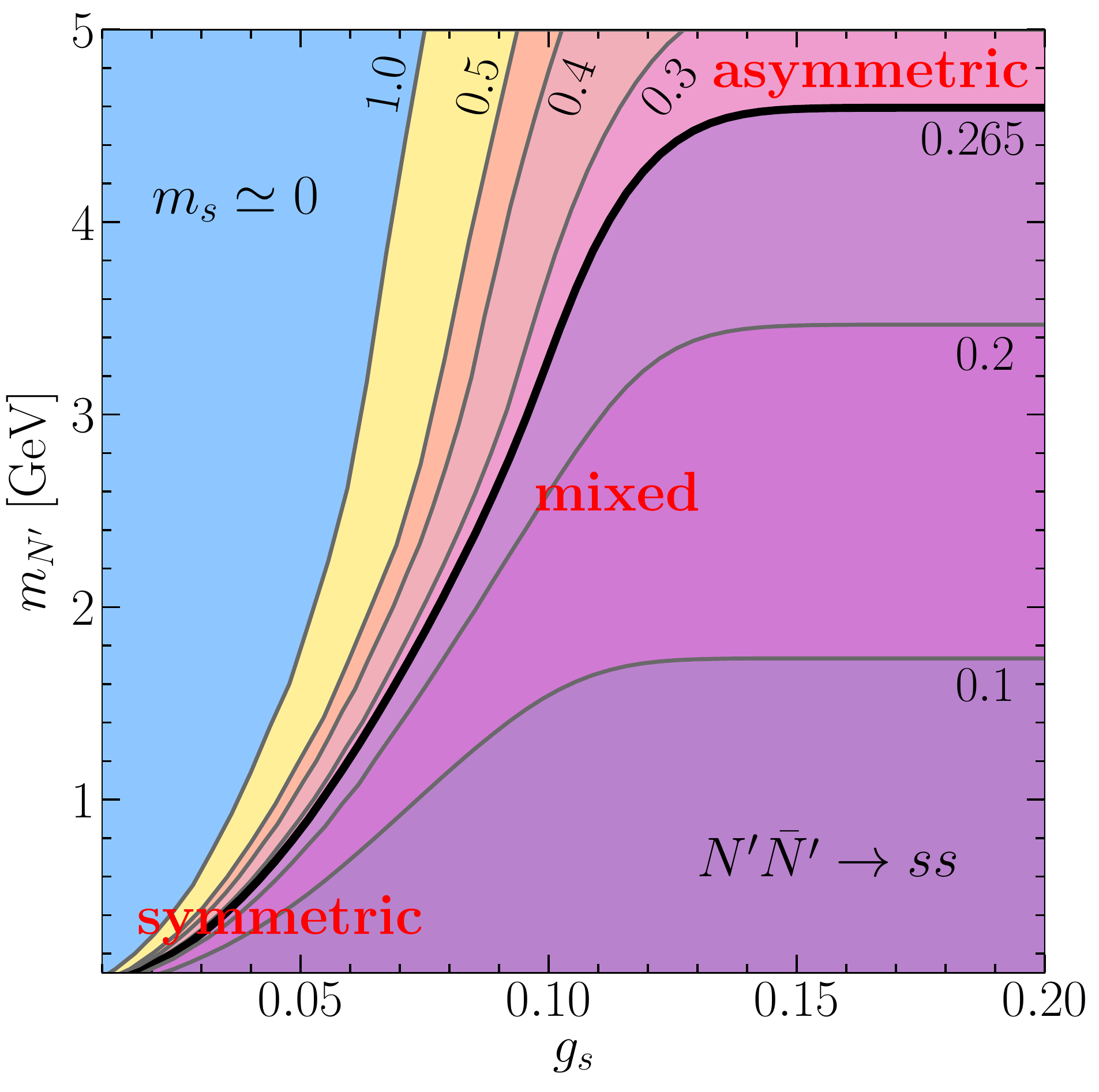}\hfill
\includegraphics[scale=0.261]{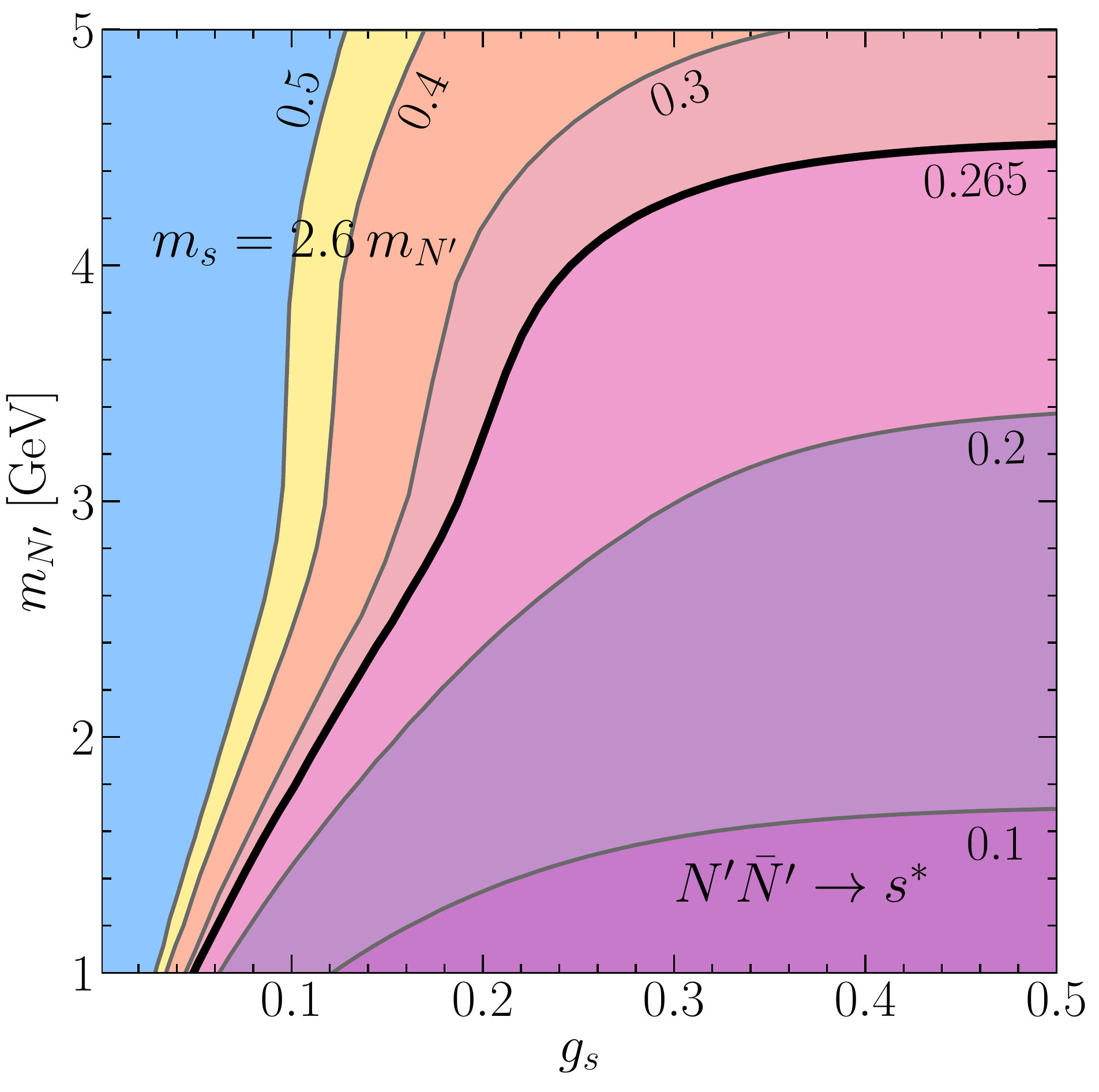}\hfill
\includegraphics[scale=0.261]{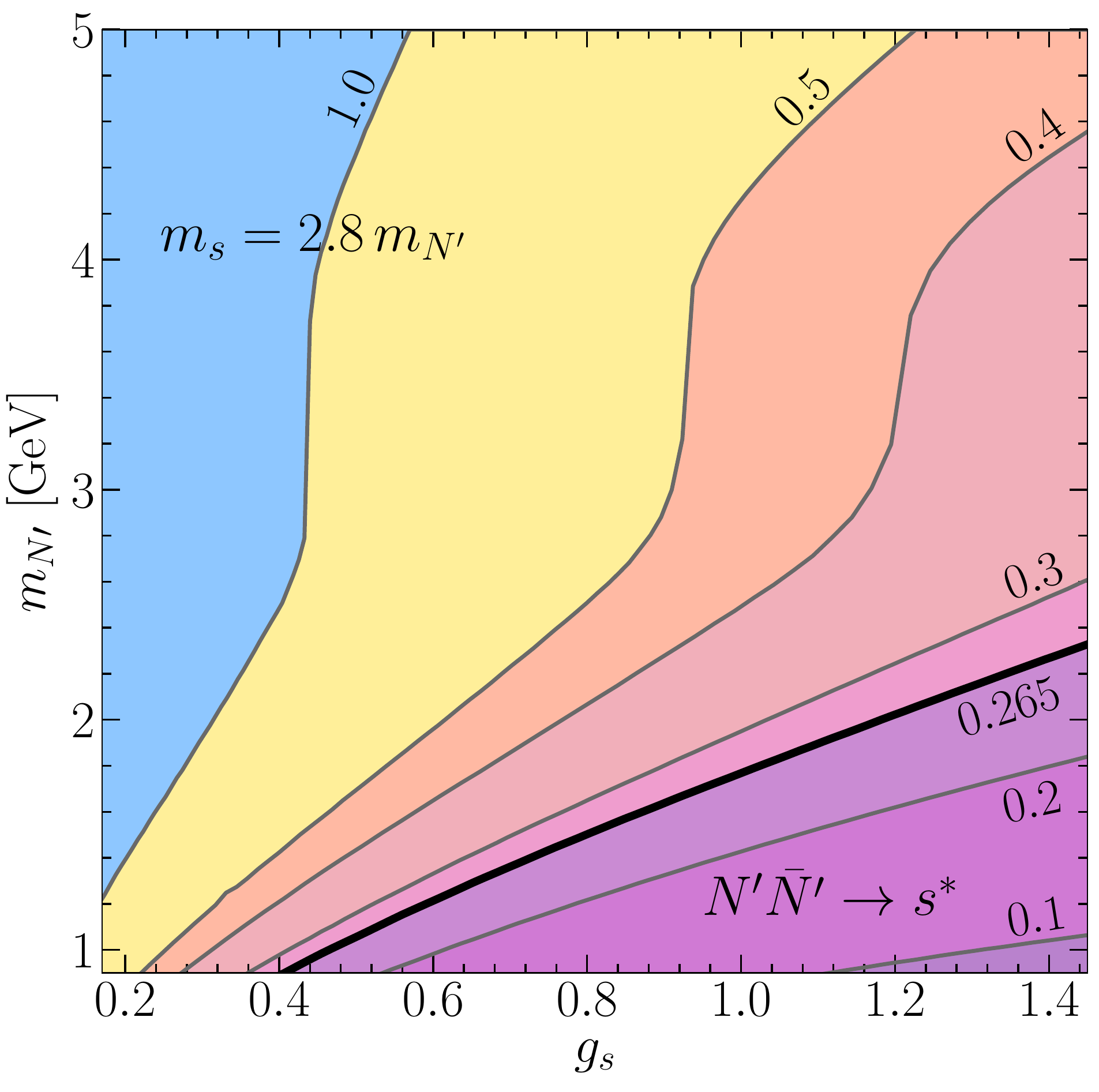}}
 \caption{Contours of DM relic density
$\Omega_{N'}\equiv \rho_{N'}/\rho_{\rm crit}$ in the plane of
DM mass versus coupling to singlet, for three relations of singlet mass $m_s$ to the DM mass
$m_{N'}$.  Left: $m_s\ll m_{N'}$, with $N'\bar N'\to ss$ annihilation.  Center: $m_s = 2.6\, m_{N'}$
with $N'\bar N'\to s^*$ (virtual $s$) annihilation.  Right: like center, but with $m_s = 2.8\, m_{N'}$
The heavy contour labeled $0.265$ corresponds
to the observed relic density.} 
 \label{fig:conts}
\end{center} 
\end{figure}

To find the relic abundance including both symmetric and asymmetric
components, one can solve the Boltzmann equation for their ratio
$r$ given in ref.\ \cite{Graesser:2011wi}, which depends upon
$\langle\sigma v\rangle$.  Then as shown there, the fractional 
contribution of $N'$ to the energy density of the universe is
\be
	\Omega_{N'} = \epsilon\, \eta_B\, m_{N'}{s\over \rho_{\rm crit}}
	\left(1+r\over 1-r\right)
\ee
where $\eta_B = 8.8\times 10^{-11}$ is the observed baryon-to-entropy
ratio, $s$ is the entropy density, and $\epsilon = 
\eta_{N'}/\eta_B = 123/112$ in our model (see below eq.\
(\ref{chemnet})).
Using ref.\ \cite{Iengo:2009ni}, we checked whether the DM 
annihilation cross section might be 
Sommerfeld-enhanced since $m_{s} < m_{N'}$, but this was
a negligible effect in the relevant parts of parameter space that we 
will specify below.
In Figure\ \ref{fig:conts} (left) we plot contours of $\Omega_{N'}$, the
fractional contribution of the DM to the energy density of the
universe, in the plane of
$m_{N'}$ versus $g_s$.  For $g_s\gtrsim 0.14$ the maximum value
in eq.\ (\ref{massbound}) is achieved, whereas for lower $g_s$, the
symmetric component abundance is increased (while the asymmetric
abundance remains fixed), corresponding to lower DM masses.

In the opposite regime $\lambda_s\gg g_s$, the annihilation
$N'\bar N'\to ss$ could
in principle be dominated by the $s$-channel diagram, giving 
the cross section
\bea
	\langle\sigma v\rangle &\cong& {1\over \pi}\left(3\lambda_s v_s
g_s\over 8\, m_{N'}^2\right)^2 \, \bar F(x),
\label{sigv}
\eea
in the case where $m_s\ll m_{N'}$, with
\be
	\bar F(x) = {x\over K_2(x)^2}\int_0^\infty d\epsilon\, 
	{\epsilon^{3/2}\over(1+\epsilon)^{5/2}} \,
	{K_1(2 x\sqrt{1+\epsilon})}
\ee
For $x\sim 20$, $F\cong 0.01$ leading to the requirement that
$g_s$ must be significantly larger than in the previous case
to suppress the symmetric DM component.  Such values are
excluded by direct DM search constraints to be discussed in section\
\ref{ddsect} below.  
Hence there is no practical enlargement of the allowed parameter space from
including the $s$-channel contribution.

\subsubsection{$N'\bar N'\to$ SM annihilation}

In the other case where $m_s > m_{N'}$, the total 
annihilation cross section for $N'\bar N'$ into $\mu^+\mu^-$, 
$\pi^+\pi^-$, {\it etc.,} through the Higgs portal, does not depend upon the
couplings of $s$ to the final state particles nor on the number of
decay channels, in the limit of the 
narrow-width approximation for the intermediate virtual $s$.  In this
limit we can approximate the Breit-Wigner distribution for the $s$ 
propagator as a $\delta$ function, $(\pi/\Gamma_s)\delta(s-m_s^2)$ [$s$
is the Mandelstam variable], and
the couplings in the singlet decay width $\Gamma_s$ cancel against
those in the annihilation amplitude.  One can think of this as the cross section
for $N'\bar N'\to s$, which one integrates over the $\delta$ function when
doing the thermal average.  In this way we find
\be
	\langle\sigma v\rangle \cong \pi {g_s^2\over 2 m_{N'}^2 }\,
{(y^2-1)^{3/2}\,x\over K_2(x)^2}\, K_1(2 x y)
\label{sigmavNWA}
\ee
where $x = m_{N'}/T$ as usual, and $y \equiv m_s/(2\, m_{N'})$.  It turns on 
steeply above the threshold $y=1$ for resonant enhancement, and then
quickly decays because of the Boltzmann suppression for $y\gg 1$.
Nevertheless we find that it can be large enough for values of $y\lesssim
1.3-1.4$ that are not finely tuned to be close to 1, as we show in Figure\
\ref{fig:conts} (center and right plots).  

We will see that for such parameter values, the
$t$-channel exchange of $s$ for $N'$ scattering on nucleons can still
be consistent with direct detection constraints.  In this process, the
suppression by the small coupling of $s$ to nucleons (through the
singlet-Higgs mixing angle $\theta_{s}$) is not canceled by anything,
in contrast to the $s$-channel resonance.

\section{Neutrino properties and HNL constraints}
\label{neutrino}
Below the scales of electroweak symmetry breaking and the
HNL mass $M_{N}$, 
the light neutrino mass matrix gets generated,
\bea
	m_\nu &\cong&  \epsilon_\nu -\delta M' \nn\\
	\delta M' &\equiv& {\bar v^2\over M_N^2}\, \eta_\nu^T\, 
	\epsilon_\nu\, 
	\eta_\nu
\label{mnu}
\eea
However $|\eta_\nu| \bar v/M_N\ll 1$ is the magnitude of the mixing between
the light neutrinos and the HNLs, as we will discuss below, so that
the correction $\delta M' \ll \epsilon_\nu$ can be ignored.  We
reiterate
that $\epsilon_\nu$ is generated by the usual seesaw mechanism,
integrating out sterile neutrinos whose mass is above all the other
relevant scales in our model.

Recall that the stability of the dark matter $N'$ requires 
$\eta_\nu$ to be a matrix with one vanishing eigenvalue, which implies
that  the lightest neutrino is massless.  This is an
exact statement, not relying upon the neglect of $\delta M'$,
since $\epsilon_\nu$ and $\delta M'$ are 
simultaneously diagonalizable by construction.  This is a consequence
of our MFV-like assumption that $\eta_\nu$ is the only source of
flavor-breaking in the HNL/neutrino sector.

\subsection{Explicit $\eta_\nu$ and HNL mixings}

Using eq.\ (\ref{epsrel}) we can solve for $\eta_\nu$ in terms of the neutrino masses and mixings,
\be
	\eta_\nu = O\,\left({D_\nu\over\bar\mu_\nu}\right)^{1/2}U_\mathrm{PMNS}^{-1}
\label{etaeq2}
\ee
where $D_\nu$ is the diagonal matrix of light $\nu$ mass eigenvalues,
and
$U_\mathrm{PMNS}$ is the $3\times3$ PMNS matrix. The orthogonal matrix
$O$ is undetermined since the $N_i$ are practically degenerate; for
simplicity we set it to $1$ in the following.
 Since we have assumed that one eigenvalue is
vanishing, the other two are known,
\bea
D_{\nu 11} &=& 0,\ D_{\nu 22} = \sqrt{\Delta m^2_{21}},\ 
	D_{\nu 33} = \sqrt{\Delta m^2_{31}}, \hbox{\quad
\qquad\qquad NH}\nn\\
	D_{\nu 33} &=& 0,\ D_{\nu 22} = \sqrt{\Delta m^2_{32}},\
	 D_{\nu 11} = \sqrt{\Delta m^2_{32}-\Delta m^2_{21}}, 
	\hbox{\quad\  IH}
\eea
for the normal and inverted hierarchies, respectively.
 
The light neutrinos mix with $N_i$, with mixing matrix elements
given by
\be
	U_{\ell i}  \cong {\eta^T_{\nu,\ell i} \bar v\over M_N}
\label{thetanu}
\ee
where $\ell = e,\mu,\tau$ and $i=1,2,3$.
Constraints on $U_{\ell i}$ arise from a variety of beam dump
experiments and rare
decay searches, summarized in refs.\
\cite{Drewes:2015iva,Alekhin:2015byh}. As we now discuss, the
applicability of these limits depends upon whether the scalar singlet
is heavier or lighter than the HNL's, since this determines the 
dominant decay modes of the latter.

\subsubsection{Unitarity constraints for $m_s < M_N$ case} 

\begin{figure}[t]
\begin{center}
\centerline{\includegraphics[scale=0.65]{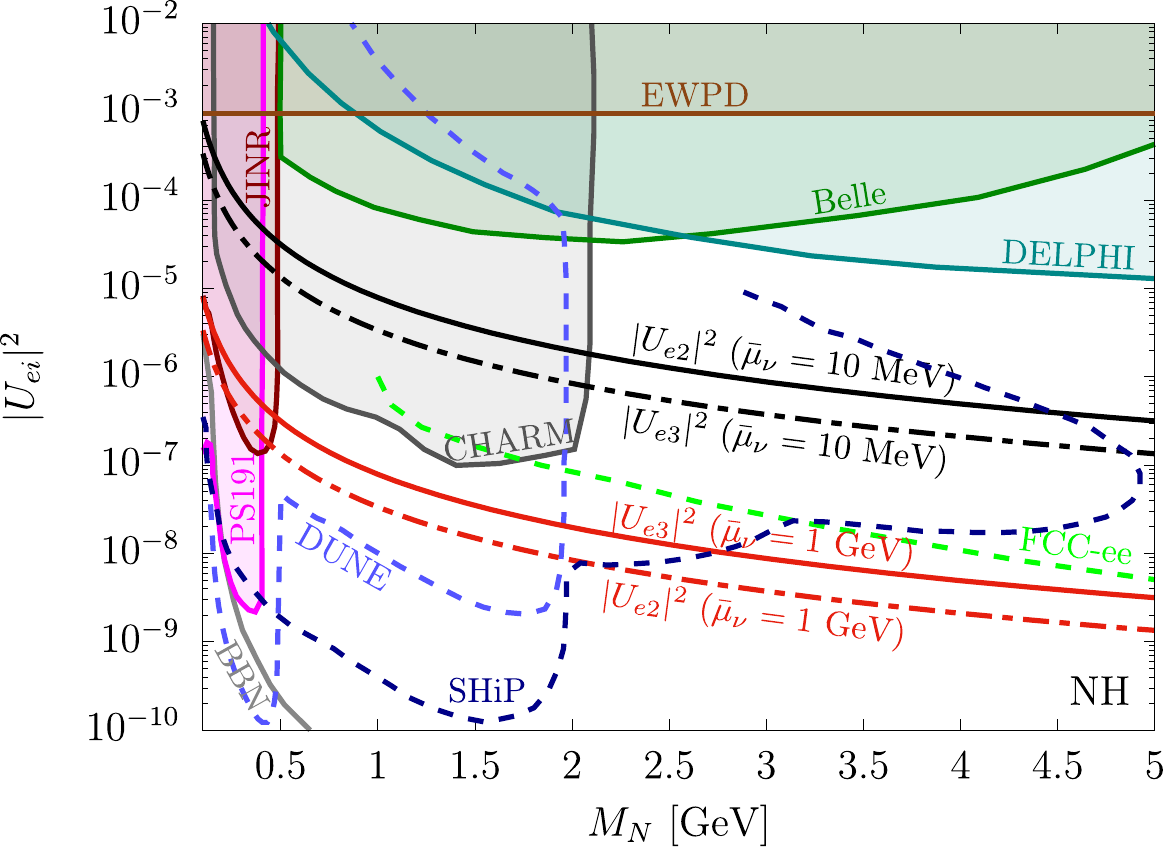}
\hfil \includegraphics[scale=0.65]{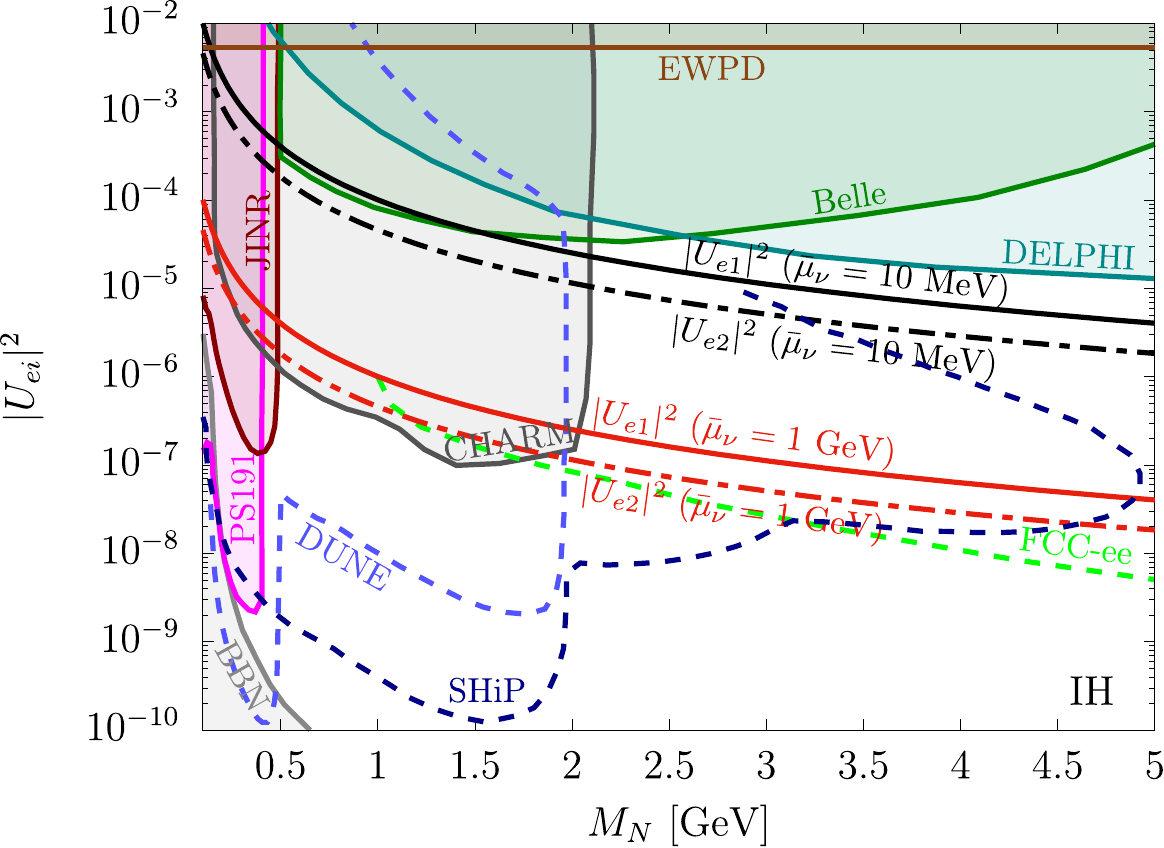}}
 \caption{Summary of constraints on HNL mixing with electron
neutrinos, over mass range of interest for our model (left: normal
 hierarchy, right: inverted hierarchy).
 Solid and dot-dashed black and red curves show the model's predictions for $U_{e2}~(U_{e1})$ (solid curves) and 
$U_{e3}~(U_{e2})$ (dot-dashed) in the normal (inverted) mass hierarchy, for two choices of the parameter $\bar\mu_\nu$ that
determines the mixing through eqs.\ (\ref{etaeq2}, \ref{thetanu}).
$U_{e1}\equiv 0~(U_{e3}\equiv0)$ for the normal (inverted) hierarchy since $N_1=N'~(N_3=N')$ denotes the dark
matter HNL.
Sensitivity regions of future experiments FCC-ee~\cite{Blondel:2014bra},
 DUNE~\cite{Krasnov:2019kdc} and SHiP~\cite{SHiP:2018xqw} are bounded 
by dashed curves.}
 \label{fig:hnls}
\end{center} 
\end{figure}

If $m_s<M_N$, then
many of the beam-dump and other limits on the mixing angles
(\ref{thetanu}) versus
$m_N$, shown in Figure\ \ref{fig:hnls},  cannot be
directly applied to our model because they assume that $N$ decays
are mediated only by the weak interactions, through $N$-$\nu$ mixing,
whereas we have a more efficient decay channel $N\to \nu s$, from the
$g_s\bar N_i N_i$ coupling and mixing.  All of the bounds that rely
upon detecting visible particles from the decay will now be sensitive
to the singlet mass $m_s$ and mixing angle $\theta_{s}$ with the Higgs,
and not just $M_N$.  To modify these limits appropriately would
require a dedicated reanalysis of each experiment, which is beyond the
scope of our work.

However  we can still make a definite 
statement about how weak the
limit on $N$-$\nu$ mixing could possibly be, even in the case where the singlet escapes the
detector unobserved, because electroweak precision data (EWPD) are 
only
sensitive to the reduction in the SM couplings caused by the mixing,
that we can readily calculate.  This is most straightforward in the
basis of the mass eigenstates, where $\eta_\nu$ is diagonal.  Then the
mass matrix (\ref{33mm}) is block diagonal, and there is a mixing
angle $\theta_i$ connecting each pair of light and heavy mass
eigenstates.  The relation between the flavor states (labeled by subscript 
$\alpha$) and the mass eigentstates (labeled by $i$) is
\be
	\nu_\alpha = (U_{\mathrm{PMNS}})_{\alpha i}\,\cos\theta_i\,\nu_i
\equiv N_{\alpha i}\,\nu_i
\label{Ni}
\ee
In refs.\ \citep{Antusch:2014woa,Fernandez-Martinez:2016lgt}, the matrix $N_{\alpha i}$ is introduced in this way to parametrize departures from unitarity in the lepton mixing matrix, and the magnitudes of $NN^\dagger$ are constrained by various precision electroweak data. The elements of such a matrix can be written in our model as
\be
	|NN^{\dagger}|_{\alpha \beta} \equiv \Big|\sum_i N_{\alpha i} N_{i \beta}^{\dagger} \Big| = \Big| \delta_{\alpha \beta} - \sum_i (U_{\mathrm{PMNS}})_{\alpha i}\, \sin^2{\theta_i}\, (U_{\mathrm{PMNS}})_{i \beta}^{\dagger} \Big|
\label{NNd}
\ee

Since most of the constraints on physical observables are often expressed in literature in terms of the Hermitian matrix $\varepsilon_{\alpha\beta}$, defined in $N = (1 - \varepsilon)\, U_{\mathrm{PMNS}}$\ \cite{Fernandez-Martinez:2016lgt}, we have that the predicted $\varepsilon_{\alpha\beta}$ turns out to be~\footnote{The matrix $\varepsilon$ defined here is called $\eta$ in ref.\ \cite{Fernandez-Martinez:2016lgt}.}
\be
	\varepsilon_{\alpha \beta} = \sfrac{1}{2}\, \Big| \sum_i (U_{\mathrm{PMNS}})_{\alpha i}\, \sin^2{\theta_i}\, (U_{\mathrm{PMNS}})_{i \beta}^{\dagger} \Big|
\label{etaab}
\ee
The most stringent limits on $\varepsilon_{\alpha\beta}$ that can be applied to our model come from the measurement of the $W$ boson mass, which depends upon the combination \cite{Fernandez-Martinez:2016lgt} 
\bea
M_W &\simeq& M_W^{\mathrm{SM}}
\left[(NN^\dagger)_{ee} (NN^\dagger)_{\mu\mu}\right]^{1/4}
 {s_\W^{\mathrm{SM}}\over s_\W}\nn\\
&\cong& M_W^{\mathrm{SM}}\, (1 + 0.20\, 
(\varepsilon_{ee} + \varepsilon_{\mu\mu}))
\label{MWcorr}
\eea
where the SM radiative corrections, 
parametrized by the variable $\Delta r = 0.03672$\ 
\cite{Tanabashi:2018oca}, are included in the computation; they enter
through the weak mixing angle \cite{Antusch:2014woa},
\be
	s^2_\W = \frac12\left[1-\sqrt{1 - {2\sqrt{2}\pi\alpha\over
G_\mu M_{Z}^2}(1+\Delta r)\left[(NN^\dagger)_{ee}
(NN^\dagger)_{\mu\mu}\right]^{1/2}}\, \right]
\ee
Using the experimental and SM values of $M_{W}$ in eq.\ (\ref{MWcorr}), 
we obtain a $95\%$ C.L. upper bound on $(\varepsilon_{ee} + \varepsilon_{\mu\mu}) \leq 2.64 \times 10^{-3}$.

In our framework, the mixing angles $\theta_{i}$ in eq.~\eqref{etaab}
and be computed explicitly, from the eigenvalues of $\eta_\nu$, up
to multiplicative factors,
\be
	\theta_i \cong {\eta_i\,\bar v\over M_N}
\label{thetaeqn}
\ee
where $\eta_i$ is the eigenvalue of $\eta_\nu$ associated with the
eigenvector that couples to $N_i$.  For the normal hierarchy, we label
$\eta_1 = 0$ for the massless state, while for inverted hierarchy 
$\eta_3=0$.
Using (\ref{etaeq2}), we can solve explicitly for $\eta_\nu$ in either mass
scheme, up to the overall proportionality controlled by the parameter
$\bar\mu_\nu$. Comparing the combination $(\varepsilon_{ee} + \varepsilon_{\mu\mu})$, computed from eqs.~\eqref{etaab} and  (\ref{thetaeqn}), to the upper limit found above from $M_{W}$, yields lower bounds on the scale $\bar\mu_\nu$ in the two mass hierarchy choices, and upper bounds on the corresponding matrices $\eta_\nu$ and the mixing angles between HNLs and
the light neutrinos.
Defining $\bar U_\ell \equiv (\sum_i |U_{\ell i}|^2)^{1/2}$, we find for the normal mass hierarchy
\bea
\label{barmunu}
	\bar\mu_\nu &>& 5.9\,\,{\rm keV} \times \left({4.5\,\,{\rm GeV}\over
M_N}\right)^2,\qquad\qquad\qquad\qquad \hbox{NH}\nn\\
	|\eta_{\nu}^T| &<& 10^{-3}\left|\begin{array}{ccc}
	0 & 0.66 & -0.32 -0.29\,i\\
	0 & 0.72 -0.05\,i & 2.14\\
	0 & -0.70 -0.04\,i & 1.91\end{array}\right|\times\left(M_N\over {4.5\,\,{\rm GeV}}\right)\nn\\
\bar U_{e} &<& 0.031,\quad  \bar U_{\mu}< 0.087,\quad
\bar U_{\tau} < 0.078\quad
\eea
The dependence on $M_N$ cancels out in the bounds
on $U_{\ell i}$.  The corresponding results for inverted hierarchy
are
\bea	
\bar\mu_\nu &>& 13.6\,\,{\rm keV} \times \left({4.5\,\,{\rm GeV}\over
M_N}\right)^2,\qquad\qquad\qquad\qquad \hbox{IH}\nn\\
|\eta_{\nu}^T| &<& 10^{-3}\left|\begin{array}{ccc}
	1.57 & 1.06 & 0\\
	-0.75-0.17\,i & 1.02 -0.12\,i & 0\\
	0.75-0.15\,i & -1.23 -0.10\,i & 0\end{array}\right|\times
\left(M_N\over {4.5\,\,{\rm GeV}}\right)\nn\\
\bar U_{e} &<& 0.073,\quad  \bar U_{\mu}< 0.050,\quad
\bar U_{\tau} < 0.056\quad
\eea
In each case the column of zeros
corresponds to the absence of coupling to the DM state $N'$; hence
we identify $N'=N_1$ for the normal hierarchy and $N'=N_3$ for the
inverted hierarchy.

We emphasize that the above bounds are robust, but might be
strengthened, depending on the choices of $m_s$ and $\theta_{s}$, 
by reanalyzing limits from other experiments to take into account the
observation of charged particles or neutral hadrons from $s$ decays following $N\to\nu
s$.  Hence the true limits are expected to lie somewhere between the
(brown) EWPD line shown in Figure\ \ref{fig:hnls} for the
normal (left) and inverted (right) hierarchy cases, and the more
stringent limits that may arise from the other (typically beam dump) experiments.

The scale $\bar\mu_\nu$ determines how the couplings 
$y_\nu = k \eta_\nu$
of the light neutrinos to the superheavy Majorana neutrinos
$\nu_R$
(as restricted
by our MFV-like assumption) are enhanced
relative to $\eta_\nu$ by
a proportionality factor, $k = ( M_{\nu_R}\bar\mu_\nu/\bar v^2)^{1/2}$.
Perturbativity of $y_\nu$ limits $k \lesssim 0.5\times 10^3$, hence the scale of
the heavy neutrinos is bounded by $M_{\nu_R}\lesssim 10^{15}\,$GeV 
for the value of $\bar\mu_{\nu}$ in eq. (\ref{barmunu}).  
This is not restrictive, and can be made consistent with our assumption
that the heavy neutrinos do not play a role during inflation or
reheating, if the reheat temperature is sufficiently low.

\subsubsection{Laboratory constraints for $m_s > M_N$ case}
If $m_s> M_N$, only three-body decays of HNL's are available, and 
they are dominated by weak interactions, induced by mixing of $N_i$ with
the light $\nu$'s.  There is also a 3-body decay $N\to\nu f\bar f$ by 
virtual $s$ exchange, but this is highly suppressed by the small scalar
mixing angle $\theta_s$ and the couplings $m_f/v$ to the Higgs.
In this case, all of the constraints on $N$-$\nu$ mixing 
shown in Figure\ \ref{fig:hnls} unambiguously 
apply.
For masses $M_{N}>2\,\,$GeV,
the most stringent limit comes from searches for  $Z\to N\nu$  
decays by the DELPHI
Collaboration~\cite{Abreu:1996pa}.  Defining again $\bar U_\ell = 
(\sum_i |U_{\ell i}|^2)^{1/2}$, 
at our largest allowed mass $M_N=4.5$\,\,GeV, the bound is 
\be
\bar U \equiv \left(\sum_{\ell}\bar U_{\ell}^2\right)^{1/2} < 0.0039\,,
\label{delphi}
\ee
since DELPHI was sensitive to the total rate of $N_i$ production from 
$Z\to N_i\nu_\ell$ decays, times the total (semi)leptonic rate of $N_i$ decays.

\begin{figure}[t]
\begin{center}
\centerline{\includegraphics[scale=0.3]{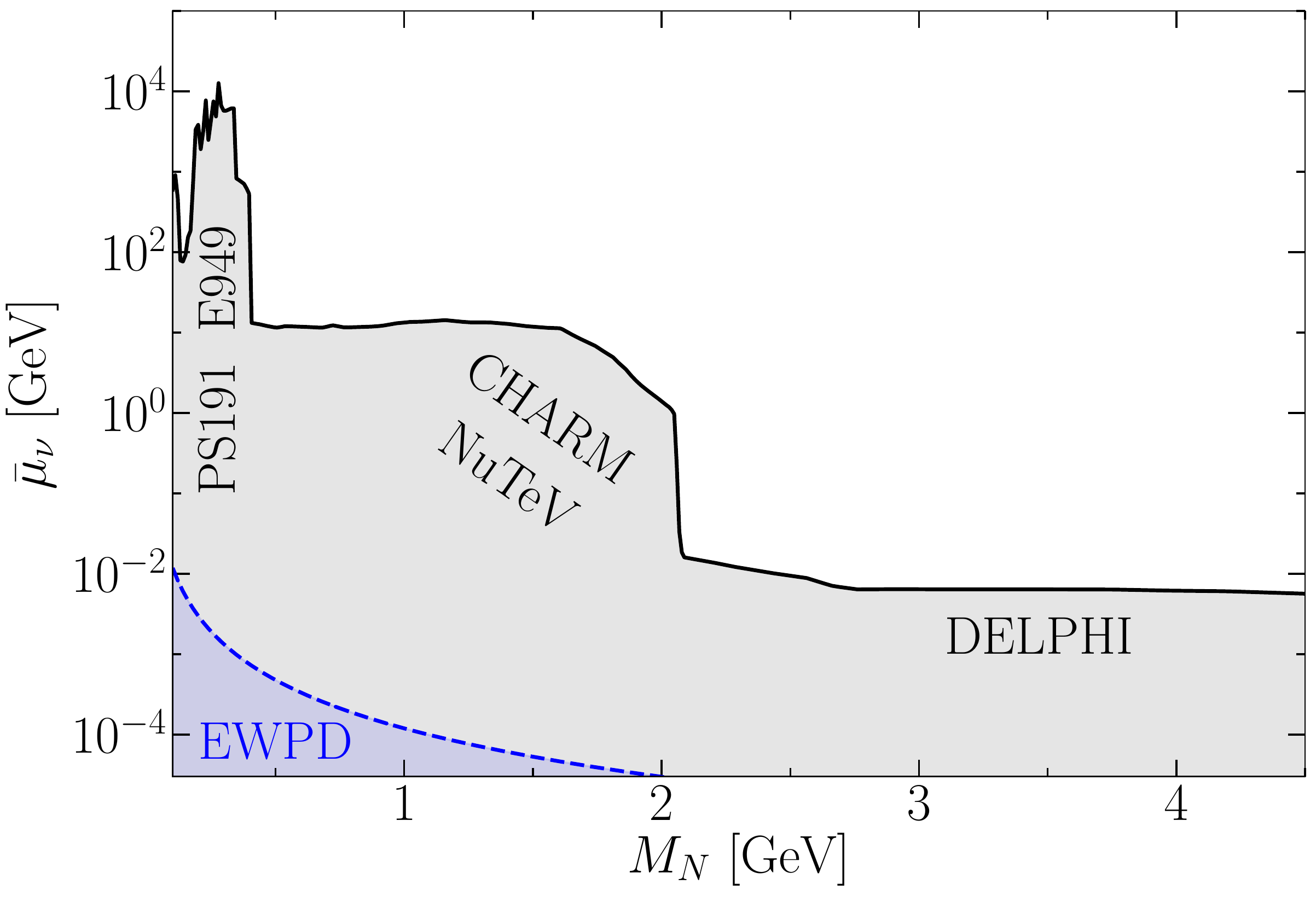}\hfil
\includegraphics[scale=0.3]{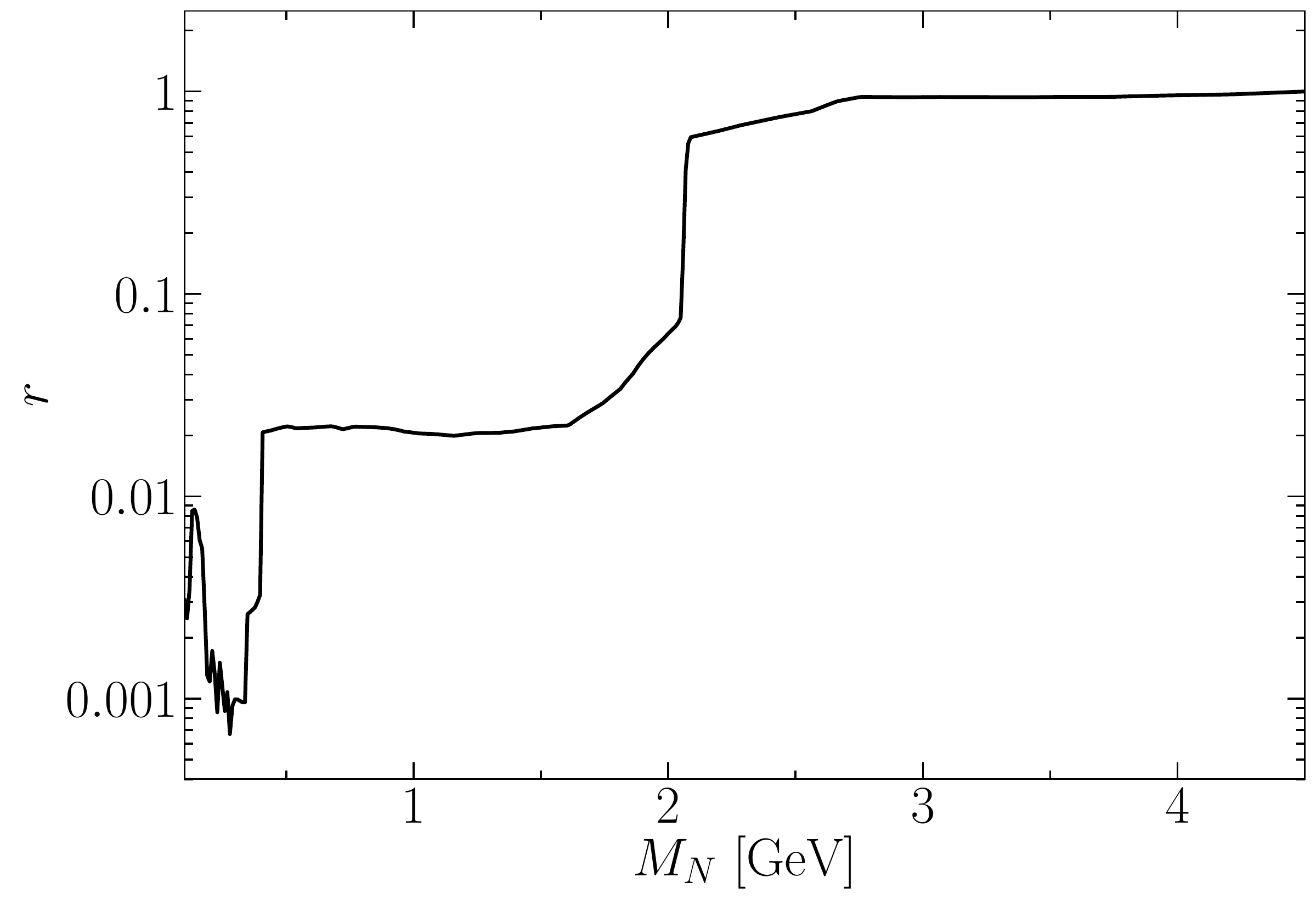}}
\caption{Left: minimum allowed mass scale $\bar\mu_{\nu} (M_N)$, predicted by our
model for the normal mass hierarchy case, 
compatible with current constraints on the HNL mixings to light 
neutrinos\ \cite{Alekhin:2015byh}. The shaded gray region is excluded.
Right: the ratio $r$ showing how the maximum allowed mixings
(\ref{mixmat}) at
$M_N = 4.5\,\,$GeV are rescaled at lower $M_N$.
}
\label{fig:rat}
\end{center} 
\end{figure}

Using eqs.\ (\ref{etaeq2}, \ref{thetanu}), the bound (\ref{delphi}) 
can be approximately saturated
if $\bar\mu_\nu = 5.7\,(9.8)\,\,$ MeV for the normal (inverted) 
hierarchy.  Taking the PDG central values 
of the neutrino masses and mixings \cite{Tanabashi:2018oca}, we find
\bea
\label{NHeta}
\eta_\nu^T &\cong& 10^{-5}\left(
\begin{array}{ccc}
 0 & 2.1         & -1.0 -0.9\, i \\
 0 & 2.3 -0.2\,i  & 6.9 \\
0 & -2.2-0.1\, i & 6.1 \\
\end{array}
\right)\nn\\
\bar U_{e} &\cong& 0.00099,\quad \bar U_{\mu }\cong 0.0028,\quad 
\bar U_{\tau } \cong 0.0025\quad
\eea
at $M_{N} = 4.5$ GeV for the normal hierarchy, and 
\bea 
\label{IHeta}
\eta_\nu^T &\cong &	10^{-5}\left(
\begin{array}{ccc}
 5.8         & 3.9   & 0 \\
 -2.8 -0.6\, i & 3.8 -0.4\,i & 0 \\
 2.8-0.6\, i  & -4.6-0.4\, i & 0 \\
\end{array}
\right)\nn\\
\bar U_{e} &\cong& 0.0027,\quad  \bar U_{\mu}\cong 0.0019,\quad
\bar U_{\tau} \cong 0.0021\quad
\eea
for the inverted hierarchy.  In each case the column of zeros
corresponds to the absence of coupling to the DM state $N'$; hence
we identify $N'=N_1$ for the normal hierarchy and $N'=N_3$ for the
inverted hierarchy.

For the lighter mass range $M_N \sim (0.4-2)\,\,$GeV, beam dump
experiments such as CHARM~\cite{Bergsma:1985is} and
NuTEV~\cite{Vaitaitis:1999wq} give the strongest limits for electron
and muon flavors, roughly $U_{ei},\,U_{\mu i}\lesssim 6\times
10^{-4}(M_N/{\rm GeV})^{-1.14}$. The largest allowed magnitudes of the
HNL mixings $U_{\ell i}$ can be expressed as a function of $M_N$,
\bea
	|U_{\ell i}| &\cong& r(M_N)
\left(
\begin{array}{ccc}
 0 & 0.00083 & 0.00054 \\
 0 & 0.00090 & 0.0027 \\
 0 & 0.00087 & 0.0024 \\
\end{array}
\right)
\label{mixmat}
\eea
focusing on the normal hierarchy case.   
We determined the minimum allowed
value of $\bar\mu_\nu$ for lower $M_N$, and the consequent
scaling factor $r(M_N)= (5.7\,{\rm MeV}/{\rm min}(\bar\mu_\nu))^{1/2}$,  
from the limits summarized in figures\ 4.10--4.12
of ref.\ \cite{Alekhin:2015byh}.  These limits 
were rescaled and combined to account for
the fact that our model has two HNLs, each of which mixes with all of the
light flavors rather than just one $N_i$ that can mix with only one
flavor at a time.
The functions ${\rm min}(\bar\mu_\nu)$ and $r(M_N)$ are plotted in 
Figure\ \ref{fig:rat}. 
The various
constraints on the HNL mixing with electron neutrinos in the mass
range relevant for our model are shown for two choices of
$\bar\mu_{\nu}$ in Figure\ \ref{fig:hnls}, including future
constraints from FCC-ee,
 DUNE and SHiP.

\subsection{$N$-$\bar N$ oscillations}

As mentioned in section\ \ref{sharing}, the $L$-violating mass term
$\delta M$ causes $N$-$\bar N$ oscillations at a rate that is too
small to destroy the lepton asymmetry in the early universe, but
fast enough to possibly be detected in laboratory searches.   In the
scenario where $m_s < M_N$, this effect cannot be observed because the
decay products of $N_i\to \nu \mu^+\mu^-$ and $\bar N_i\to \bar\nu \mu^+\mu^-$ 
differ only by having a neutrino versus antineutrino in the final states.  However
if $m_s> M_N$, the situation is more interesting.  
For the values of $\bar\mu_\nu$ and $\eta_\nu$ in eq.\ (\ref{NHeta}),
the largest eigenvalue of $\delta M$ is given by~\footnote{The eigenvalue of $\delta M$ computed in eq.\ (\ref{deltaM}) is the maximum value allowed by current experimental constraints because $\delta M \propto \bar\mu_{\nu}^{-1}$ from eqs.\ (\ref{dMeq}, \ref{etaeq2}) and the minimum of $\bar\mu_{\nu}$ is reached at $M_{N} = 4.5$ GeV as shown in Figure\ \ref{fig:rat}.}
\be
	\delta M = 3.1\times 10^{-6}{\rm\, eV}\left(2\,\,{\rm GeV}\over
	M_N\right)^2\,,
\label{deltaM}
\ee
It was recently shown by ref.\ \cite{Tastet:2019nqj} that this is 
a promising value for inducing observable $N$-$\bar N$ oscillations at the
SHiP experiment.  These
would be seen by production of $N \ell^+$ in a hadronic collision,
followed by semileptonic decays $N\to\bar N\to \ell^+ \pi$ (where $\pi$
represents a generic hadron).  The smoking gun is the presence of
like-sign leptons in the decay chain, that can only occur if $N$
oscillates to $\bar N$ within the detector. 

\subsection{Weak HNL decays}

\begin{figure*}[t]
\begin{center}
\centerline{\includegraphics[scale=0.3]{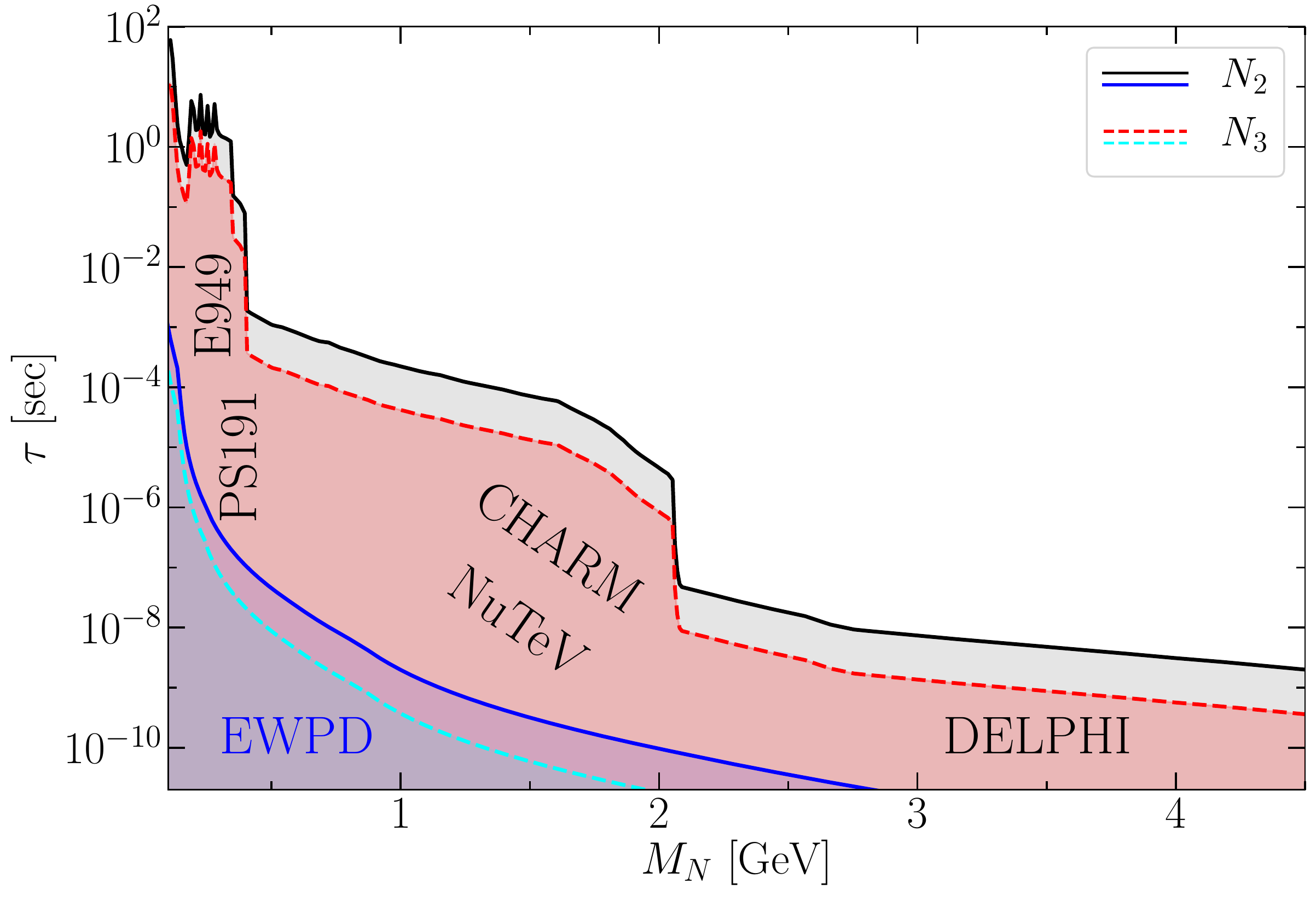}
\hfil \includegraphics[scale=0.295]{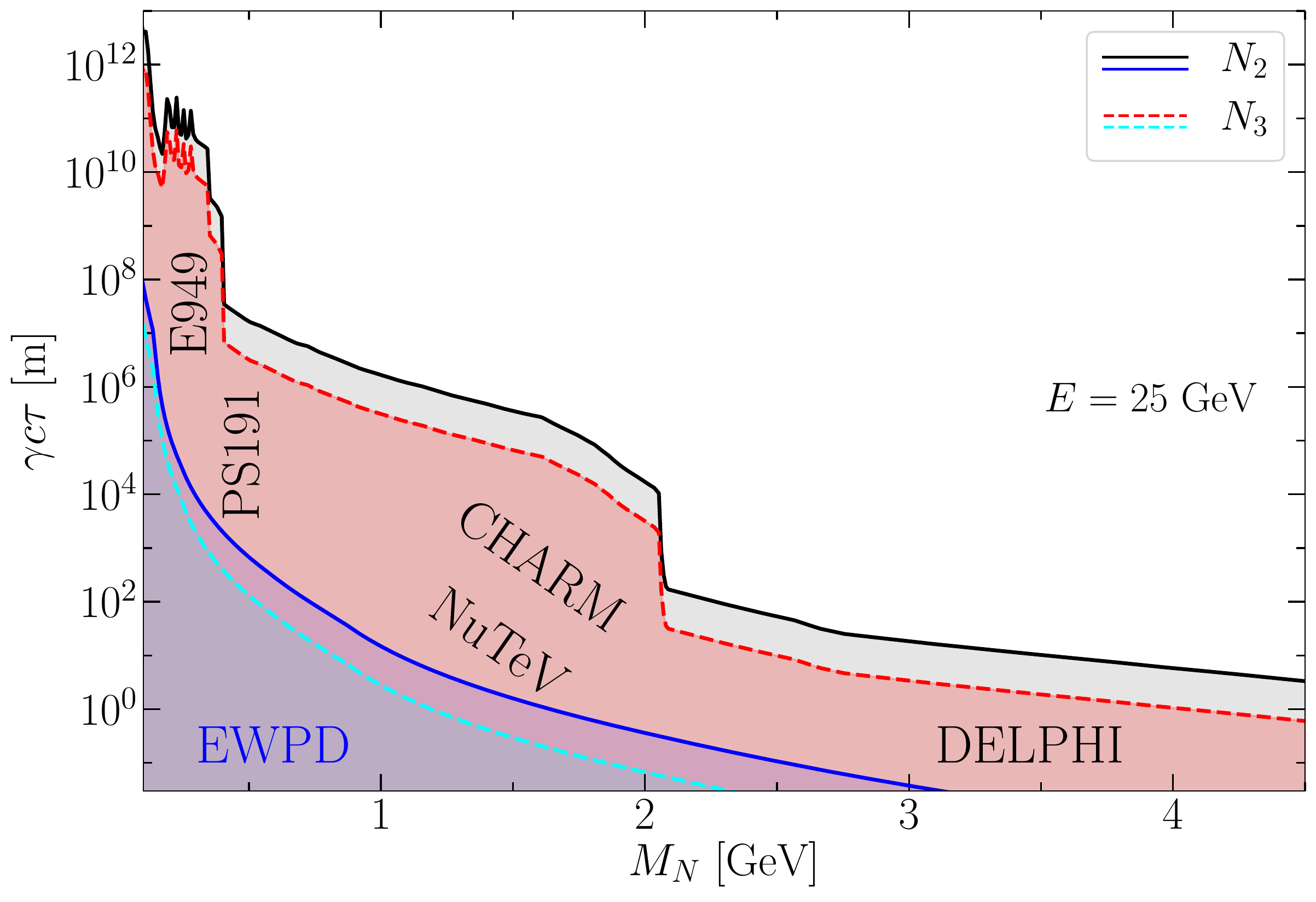}}
\centerline{\includegraphics[scale=0.3]{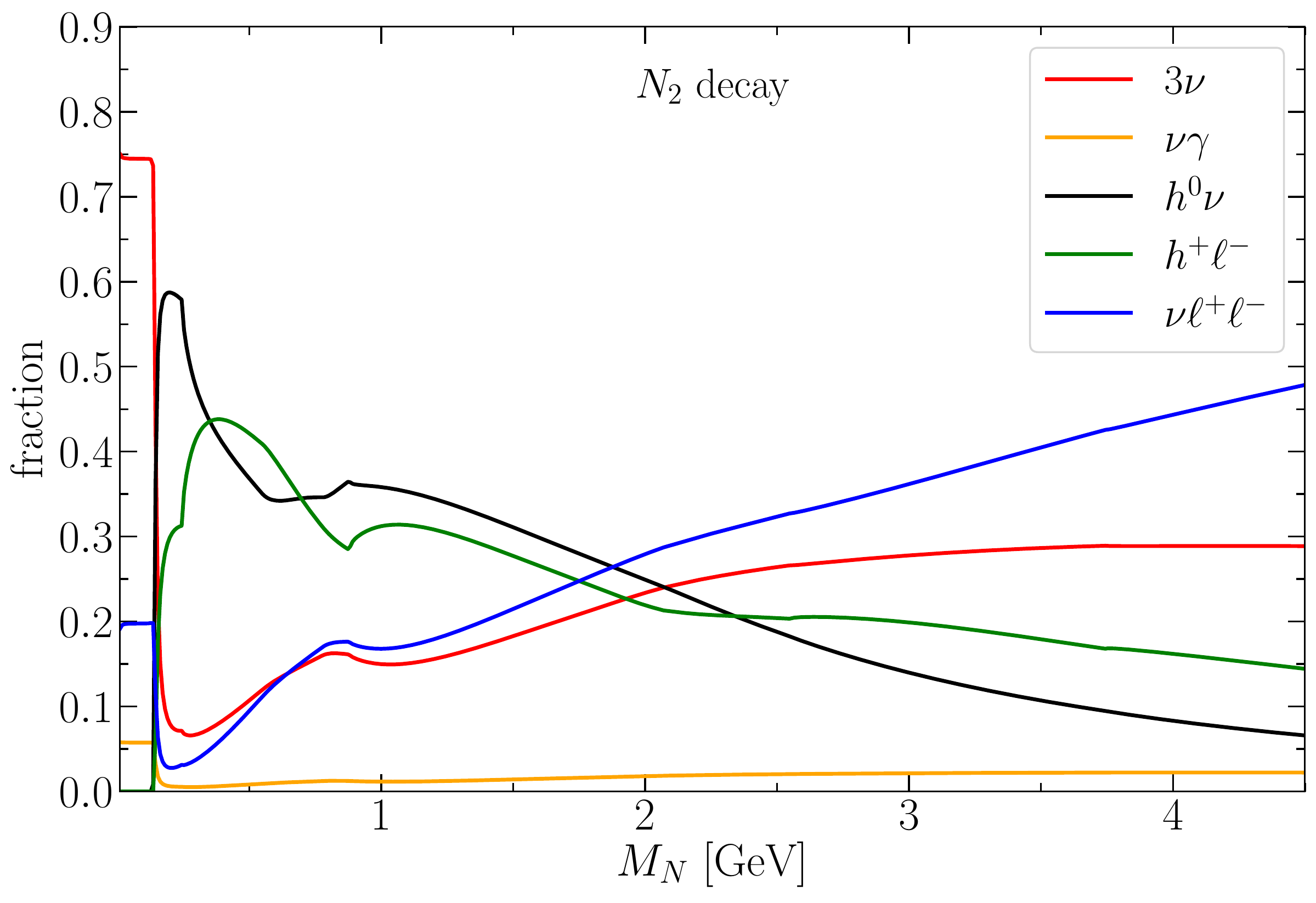}
\hfil \includegraphics[scale=0.3]{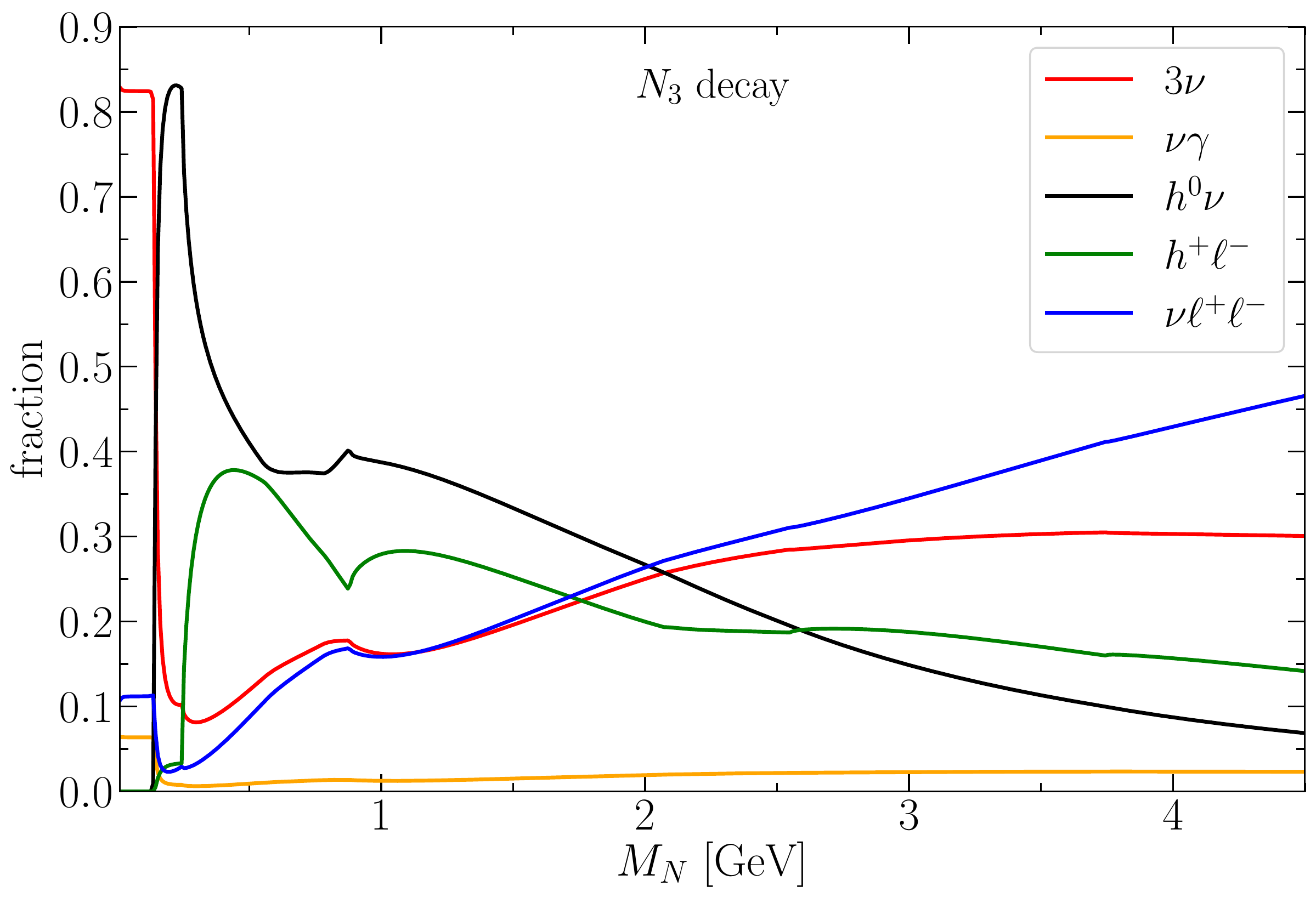}}
 \caption{Top: Minimum lifetime (left) and decay length (right) of the HNLs
$N_2$ and $N_3$, for the case of normal mass hierarchy.  Upper curves are for mass $M_N < m_s$, lower curves for $M_N > m_s$,
which determines whether weak decays or $N\to \nu\, s$ dominates.
Decay length assumes energy $E=25\,$GeV, appropriate for SHiP experiment. The shaded regions are excluded.
(The wiggles in the mass range $0.2 < M_{N} < 0.4$ GeV come 
from the E949 bound\ \cite{Artamonov:2014urb} present in figure\ 4.11
of ref.\ \cite{Alekhin:2015byh}, which also appear in
Figure\ \ref{fig:rat}.)
Bottom: branching ratios for $N_2$ (left) and $N_3$ (right) into various final states involving photon, hadrons, light neutrinos or charged leptons, for the case of weak decays, namely $M_N < m_s$.}
 \label{fig:decays}
\end{center} 
\end{figure*}

In the case where $m_s> M_N$ so that $N_i\to\nu s$ decays are blocked,
the lifetime of the unstable $N_i$ leptons is determined by weak
decays. These can be 2- or 3-body,  $N_i\to\ell^- q\bar q$ (with $q\bar q$ 
hadronizing into a meson) and 
$N_i\to \nu \ell^+\ell^-$
by $W$ and $Z$ exchange, due to mixing of $N_i$ with the active neutrinos with
mixing angles $U^T_{i\ell} \cong -U_{\ell i}$.   
Then the decay rate is of order
\be
	\Gamma_{N_i} \sim \sum_\ell {|U_{\ell i}|^2 G_F^2 M_N^5\over 192\,\pi^3}
\ee
This gives a lifetime of $\sim 10^{-3}-10^{-4}$\,s for $M_N\sim 1\,$GeV, 
making such $N_i$ decays harmless for big bang nucleosynthesis (BBN) or the CMB.  

More quantitatively, we have evaluated the partial widths for $N_i\to
\nu\gamma$, $N_i\to h^0\nu$, $N_i\to h^+\ell^-$,  $N_i\to 3\nu$,
$N_i\to \nu\ell^+\ell^-$, including the hadronic final states with
$h^0=\pi^0,\eta,\eta',\rho^0$,  $h^+ = \pi^+,K^+,\rho^+,D^+$ as
computed in ref.\   \cite{Johnson:1997cj} and\ 
\cite{Pal:1981rm}.\footnote{The formula for the decay width of $N_i\to \nu\ell_1^+\ell_2^-$
found in the literature (see refs. \cite{Johnson:1997cj, Gorbunov:2007ak})
assumes that $m_{\ell_1}$ is negligible compared to $m_{\ell_2}$.
This is not as good an approximation for the case 
$\ell_1=\mu$, $\ell_2=\tau$ as for $\ell_1=e$, $\ell_2=\mu$.
We provide the exact formula in Appendix\
\ref{appA}.}\ \  Focusing on the normal hierarchy case, we use the mixing
matrix  given by eq.\ (\ref{thetanu}) with $\bar\mu_\nu$ shown in
Figure\ \ref{fig:rat}, that leads to different lifetimes for the two
unstable HNLs $N_2$ and $N_3$.   The lifetimes are plotted in Figure\
\ref{fig:decays}, along with decay lengths in the case of HNLs with
energy $E = 25\,$GeV that would be relevant for the SHiP experiment. 
For $M_N \lesssim 0.3\,$GeV, the lifetimes start to exceed 1\,s, which
for generic models of HNLs would come into conflict with
nucleosynthesis.  In our model, this need not be the case since the
HNL abundance is suppressed by $N_i\bar N_i\to ss$ annihilations. Then
it is the singlet that should decay before BBN, which generally occurs
as long as $m_s > 2\, m_e$.  

In Figure\ \ref{fig:decays} the branching ratios for $N_2$ and $N_3$ to
decay into the various final states (summing over flavors within each
category) is also shown. Leptonic final states dominate for $M_N >
2$\,\,GeV, while hadronic ones dominate for lower $M_N$.

\subsection{Entropy and energy injection by late $N$ decays}

If the Dirac HNLs $N_2$ and $N_3$ dominate the energy
density of the universe and are sufficiently long-lived,
which may happen if the two-body decay $N_{2,3}\to\nu s$ is kinematically forbidden, 
a large amount of entropy may be injected by the decay of these particles after freezeout~\cite{Bezrukov:2009th}.
As a result, the produced dark matter relic abundance and baryon asymmetry can be diluted.   Moreover one should take
care that injected hadronic and electromagnetic energy does not disrupt the products of BBN.

In our model, since the freezeout of $N_2$ and $N_3$ occurs when they are
nonrelativistic, the number density of these particles are highly suppressed. Therefore, the entropy and energy produced by the $N_{2,3}$ decays is negligible in terms of its cosmological impact. 
We illustrate this with an example; consider $M_{N} = 1\,$GeV,
and take the freezout temperature to be $T_f\sim M_{N}/20\sim
0.05\,$GeV, for which the number of degrees of freedom in the plasma
is $g_*\cong 10$, and the decay rate is $\Gamma = 0.01\,$s$^{-1}$.
The thermal number density of the HNLs at $T = T_f$ is
$n_{N} = 4\,(M_N T_f/(2\pi))^{3/2}\,\exp(-M_N/T_f)$, and its ratio to the
entropy density at decoupling is denoted by $r_N = n_N/s$.  Then 
ref.\ \cite{Bezrukov:2009th} shows that the entropy
injected by HNL decays in this case is
\bea
     S \cong \left( 1 + 3\left(2\pi^2\,g_*\over 45\right)^{1/3}
	{(r_N\,m_N)^{4/3}\over(M_P\,\Gamma)^{2/3}}\right)^{3/4} 
	\cong  1+ 4\times 10^{-9}
\eea
where $S = 1$ corresponds to the limit of no entropy production.
This example shows that even when the lifetime is much longer than 
1\,s, the abundance is too small to create any cosmological problem. 

Previous studies show that even for decays as late as 100\,s,
GeV-scale particles are only weakly constrained by BBN.
Ref.\  \cite{Coffey:2020oir}
recently studied BBN constraints on models with late-decaying light
particles, of mass up to $1$\,GeV.  It shows that there are no constraints on electromagnetic injection for lifetimes
less than $10^4\,$s, since nuclear photodissociation processes are suppressed at earlier times.  Similarly,
ref.\ \cite{Pospelov:2010cw} finds no significant bounds
from hadronic injections for GeV-scale particles decaying 
earlier than 100\,s.

\begin{figure}[t]
\begin{center}
\includegraphics[scale=1.0]{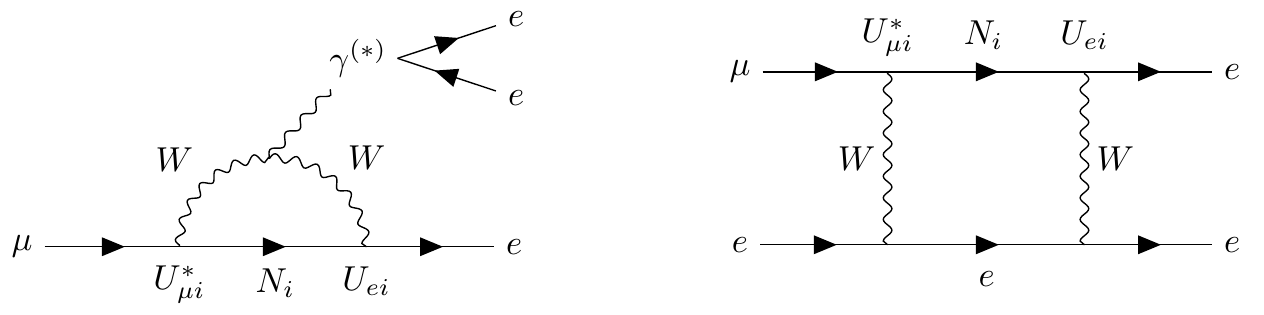} 
 \caption{Diagrams leading to $\mu\to e\gamma$ and $\mu \to 3e$
 from mixing of HNL's with the light neutrinos.
 } 
 \label{fig:lfv}
\end{center} 
\end{figure}

\subsection{Lepton flavor violation bounds}
The mixing between light neutrinos and HNL's can lead to rare lepton-flavor violating processes, analogous to the well-studied case where TeV-scale $\nu_R$'s are responsible for the seesaw mass generation \cite{lfv-constraints}.  The decays $\mu\to e\gamma$ and $\mu \to 3e$ are induced by the digrams shown in Figure\ \ref{fig:lfv}.  The most constraining process currently is
$\mu\to e\gamma$, which has a branching ratio of
\cite{Calibbi:2017uvl}
\be
\hbox{BR}(\mu\to e\gamma) = {3\,\alpha_{\text{em}}\over 32\pi}\left|
    \sum_i U_{\mu i} U_{e i}^*\, {M_N^2\over M_W^2}\right|^2 =
    {3\,\alpha_{\text{em}}\over 32\pi}\,
    |N N^{\dagger}|^2_{\mu e}\, {M_N^4\over M_W^4}
\ee
where $\alpha_{\text{em}}$ is the electromagnetic fine structure constant and $|N N^{\dagger}|_{\mu e} \equiv |\sum_i N_{\mu i} N_{i e}^{\dagger}|$ with $N_{\alpha i}$ defined in eq.\ \eqref{Ni}.
For the case of normal neutrino mass hierarchy, our least restrictive
bound based on EWPD, eq. (\ref{barmunu}), leads to 
$|\sum_i U_{\mu i} U_{ei}^*|< 1\times 10^{-3}$, and the prediction that $\hbox{BR}(\mu\to e\gamma) < 2.2\times 10^{-15}$.  This is still well below the current experimental bound of $4.2\times 10^{-13}$ set by the MEG experiment\ \cite{TheMEG:2016wtm}.

From ref.\ \cite{Dinh:2012bp} we find the branching ratio for
$\mu\to 3e$ in terms of $x \equiv M_N^2/M_W^2 \ll 1$,
\bea\hbox{BR}(\mu\to 3e)
     &\cong& {\alpha_{\text{em}}^2\over 16\pi^2\,\sin^4\theta_W}\left|\sum_i U_{\mu i} U_{ei}^*\right|^2
    x^2\left(0.6\,\ln^2 x - 0.2\,\ln x + 2.2\right)\nn\\
    &<& 1.6\times 10^{-15}
\eea
where the second line is the prediction using the value 
$|\sum_i U_{\mu i} U_{ei}^*|< 1\times 10^{-3}$ mentioned before.
 The experimental limit $\hbox{BR}(\mu\to 3e)< 1\times 10^{-12}$, set by the SINDRUM experiment\ \cite{Bellgardt:1987du}, is weaker than that of the radiative decay.

Although the lepton-flavor-violating processes currently do not constrain the model better than EWPD constraints, experimental improvements could change this in the coming years.  For example the Mu3e experiment may eventually probe $\mu\to 3e$ down to the level of $10^{-16}$ branching ratio \cite{Perrevoort:2018cqi}.
Moreover the process of $\mu N\to e N$ conversion in nuclei is expected to yield interesting limits in upcoming experiments, including Mu2e \cite{Baldini:2019elc} at Fermilab and COMET \cite{Shoukavy:2019ydh} at KEK.

 \begin{figure*}[t]
\begin{center}
\centerline{\includegraphics[scale=0.305]{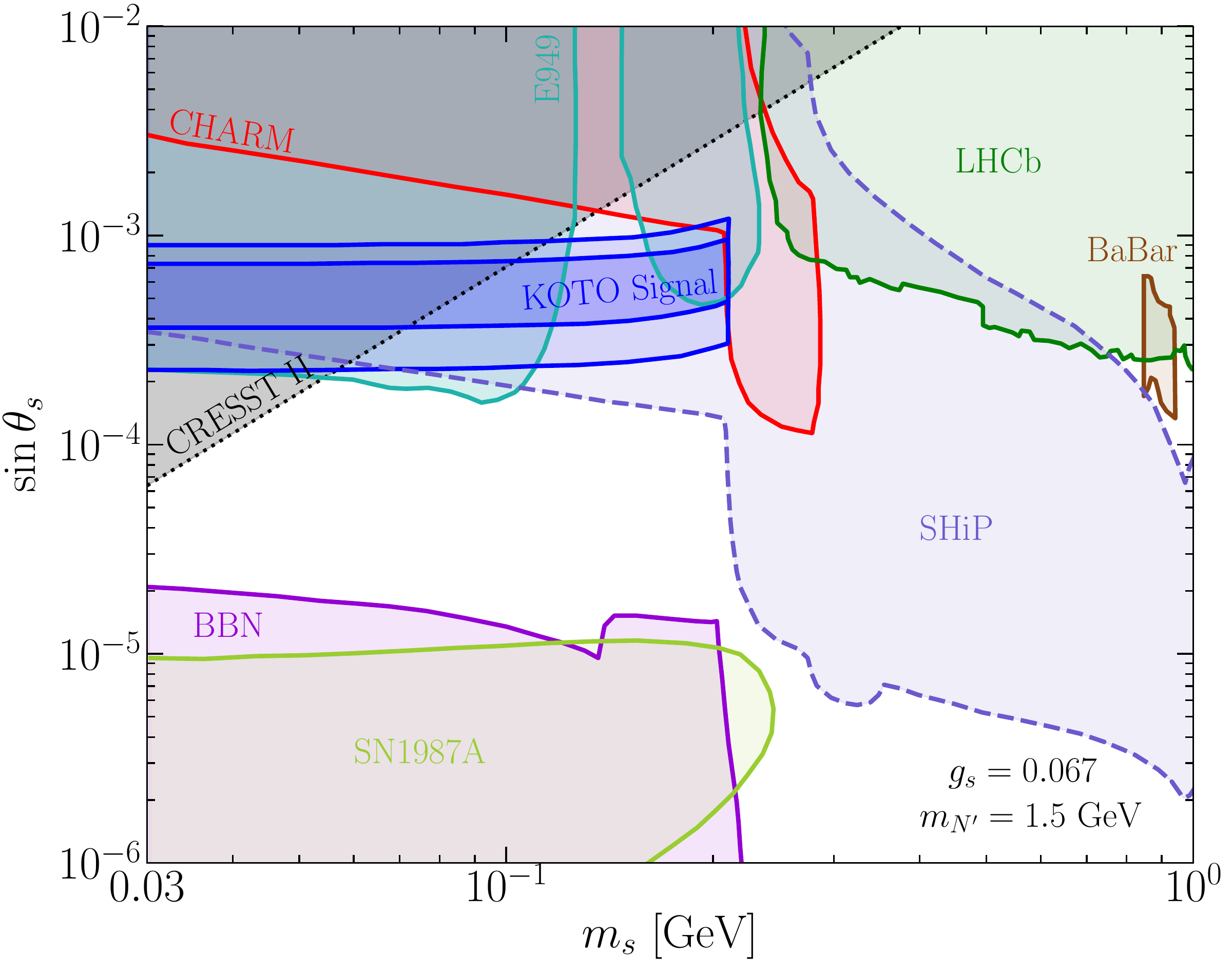} 
\includegraphics[scale=0.305]{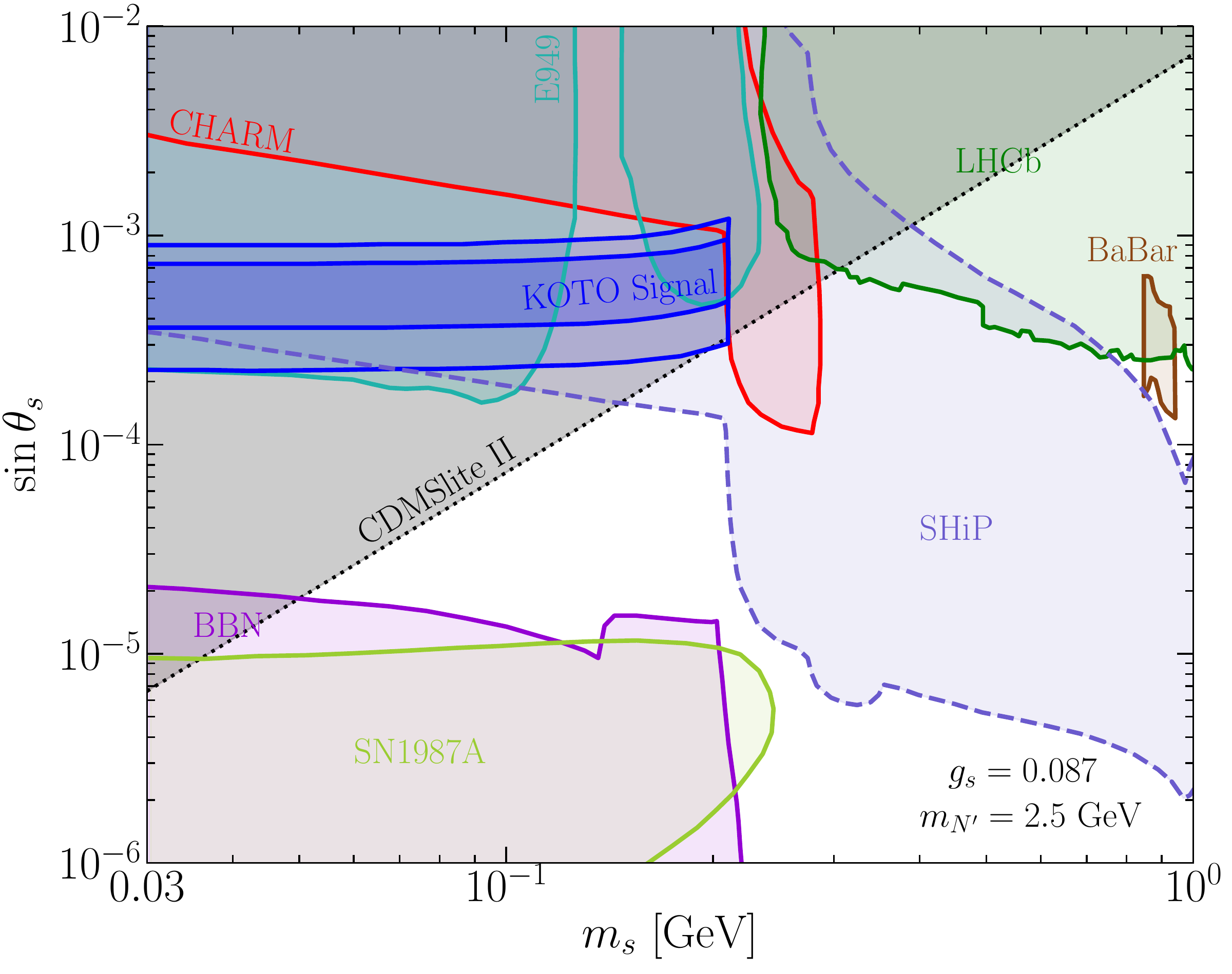}}
\centerline{\includegraphics[scale=0.305]{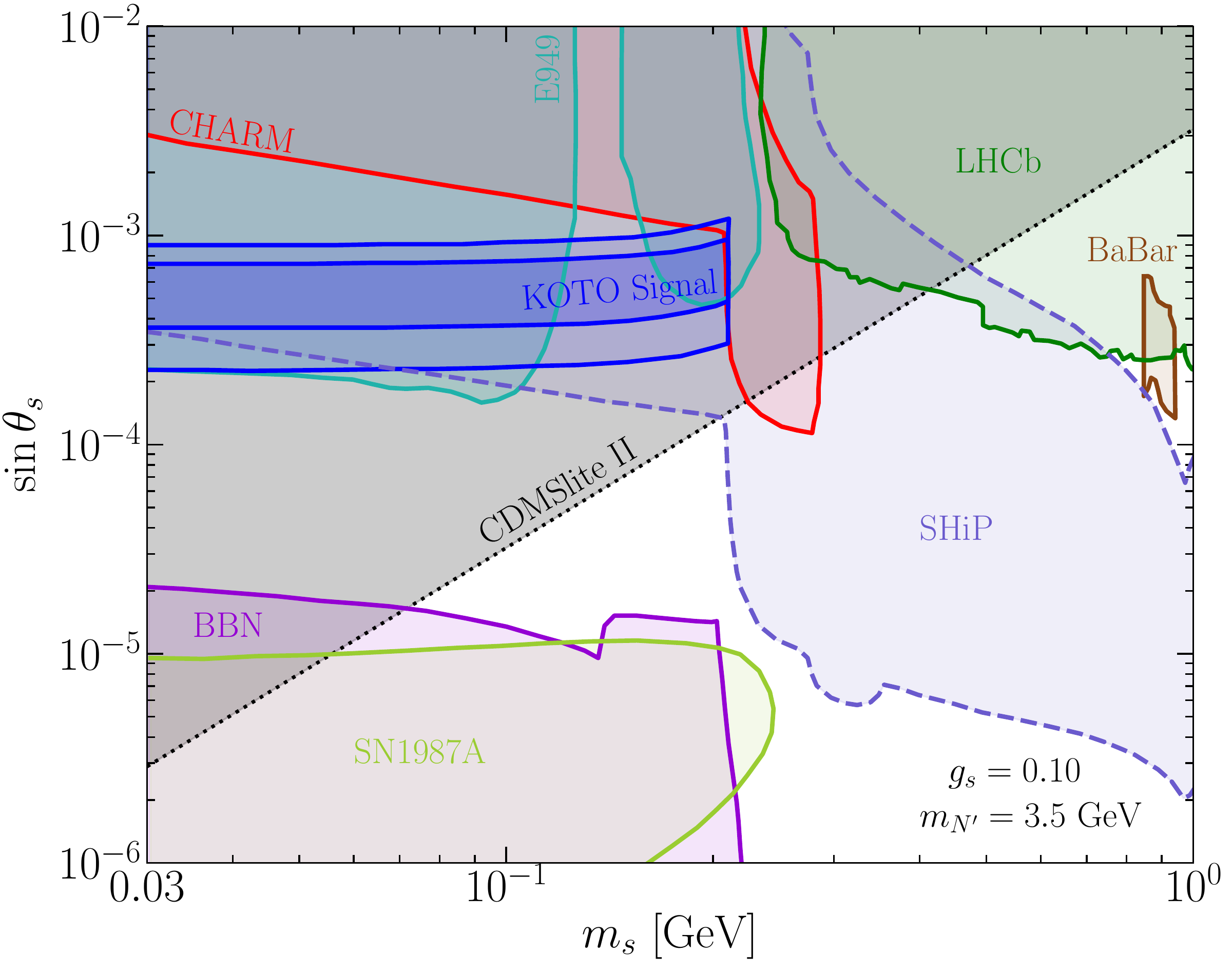} 
\includegraphics[scale=0.305]{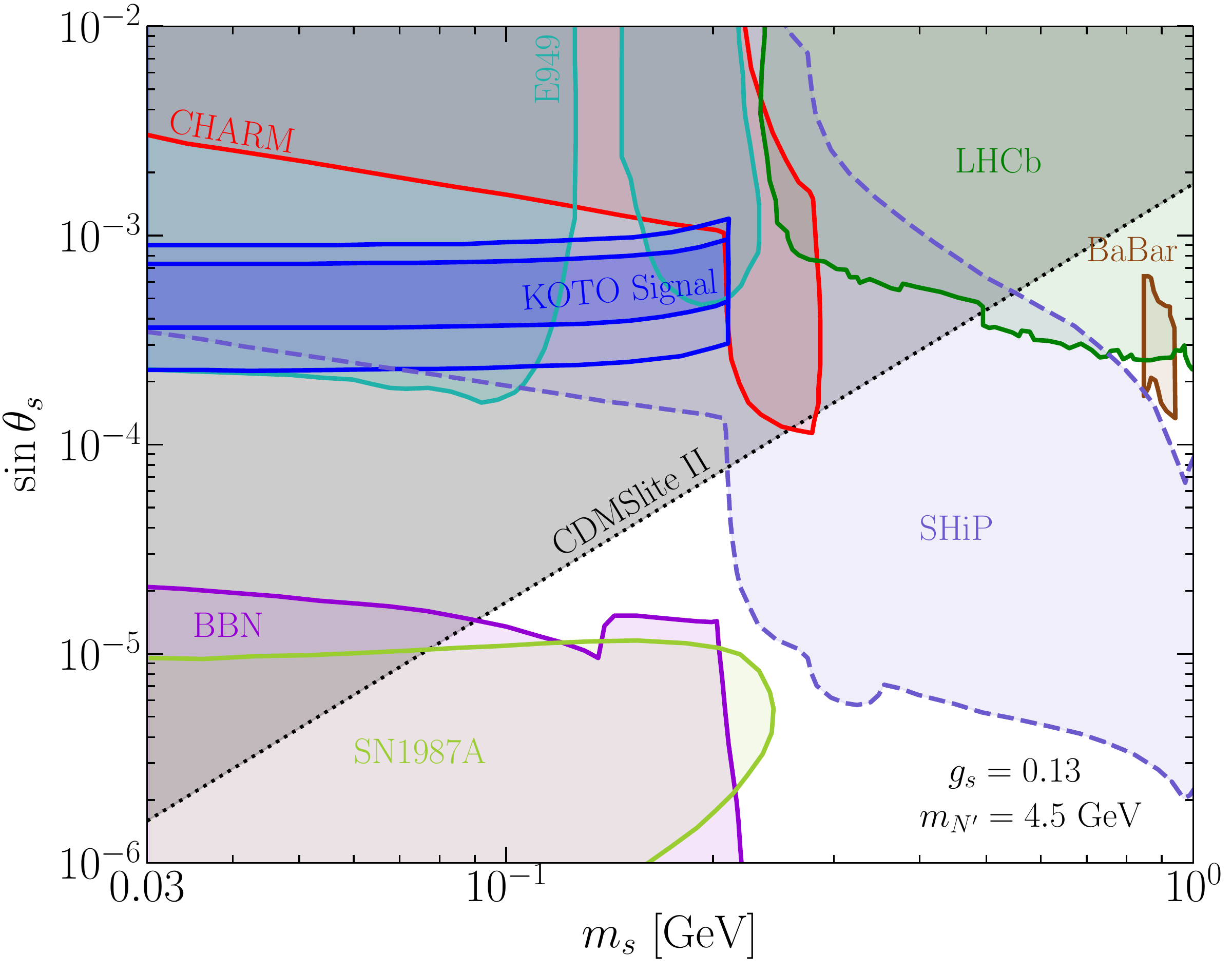}}
 \caption{Constraints on a light singlet mediator, in the
$m_s$-$\theta_s$ plane for the case $m_s < m_{N'}$.  The four plots consider different values of
the DM mass $m_{N'}= 1.5,$ $2.5$, $3.5$, $4.5$\,GeV, for which the direct
detection constraints (black dotted line) differ; all other
constraints are the same.
 The dark blue regions are favored at 1\,$\sigma$ and 2\,$\sigma$ 
 for the KOTO anomaly.
 The red, cyan, green and brown regions are excluded by CHARM\ \cite{Bergsma:1985qz}, E949\ \cite{Artamonov:2009sz}, LHCb\ \cite{Aaij:2012vr} and BaBar experiments\ \cite{Lees:2015rxq}, respectively. The violet and light-green regions are excluded by BBN\ \cite{Fradette:2017sdd} and supernova data\ \cite{Winkler:2018qyg}. Sensitivity projection for the SHiP experiment is indicated by the dashed blue-gray boundary.
 The experimental bounds, along with the projected sensitivity, are taken from
 ref.\ \cite{Winkler:2018qyg}.
 } 
 \label{fig:koto}
\end{center} 
\end{figure*}

\section{Constraints on the singlet}
\label{singlet-const}

In recent years there have been intensive efforts to constrain the
possible existence of light mediators connecting the SM to a hidden
sector, the scalar singlet with Higgs portal being a prime example.
The parameter space of $m_s$ and $\theta_s$ (the singlet-Higgs mixing
angle) is constrained by a variety of beam-dump, collider and rare
decay experiments, and by cosmology (big bang nucleosynthesis), astrophysics (supernova
cooling) and dark matter direct searches.  A large region of 
parameter space with $\theta_s\lesssim 10^{-3}$ and $m_s\lesssim
10\,$GeV remains open, and parts of this will be targeted by the
upcoming SHiP experiment \cite{Alekhin:2015byh}.  In Figure\ \ref{fig:koto} we
show some of the existing constraints, reproduced from 
ref.\ \cite{Winkler:2018qyg}.  

The KOTO experiment has searched for the rare decay 
$K_L\to\pi^0\nu\nu$ and set a new stringent upper limit of $3\times
10^{-9}$ on its branching ratio 
\cite{Hutcheson:2019qfe,Beckford:2019cuc}.  Recently four candidate 
events above expected backgrounds  were reported \cite{Shinohara}, far
in excess of the standard model prediction (BR = $3\times 10^{-11}$\ \cite{Tanabashi:2018oca}). 
These could potentially be explained by the exotic decay mode
$K_L\to\pi^0 s$, if $s$ is sufficiently long-lived to escape
detection, or if it decays invisibly.  Such an interpretation has been
previously considered in refs.\
\citep{Kitahara:2019lws,Egana-Ugrinovic:2019wzj,Dev:2019hho,Liu:2020qgx}.  In Figure\ \ref{fig:koto},
the 1$\sigma$ and 2$\sigma$ regions estimated in ref.\
\cite{Egana-Ugrinovic:2019wzj} for explaining the KOTO
excess are shown in blue.  Parts of these regions are excluded by
other experiments, notably NA62~\cite{Ruggiero} and E949~\cite{Artamonov:2009sz}, but a range around
$m_s \sim (120-160)$\,MeV and $\theta_s\sim(2-9)\times 10^{-4}$ remains
viable.

The four plots in Figure\ \ref{fig:koto} pertain to different choices of
the DM mass $m_{N'}$, for the purpose of showing constraints from 
direct detection, that we describe in the following section.  It can
be seen that the region favored by the KOTO excess events is excluded
by DM direct searches except for light DM, with $m_{N'}\lesssim
2.5\,$GeV.

\section{DM direct/indirect detection and self-interactions}
\label{ddsect}
In general, the interactions of DM with nucleons versus with other
particles are independent processes, whose cross sections need
not be related.  In our model however, both are mediated by exchange
of the singlet $s$, so it is natural to consider them together.

\subsection{DM-nucleon scattering}

\begin{figure*}[t]
\begin{center}
\includegraphics[scale=0.4]{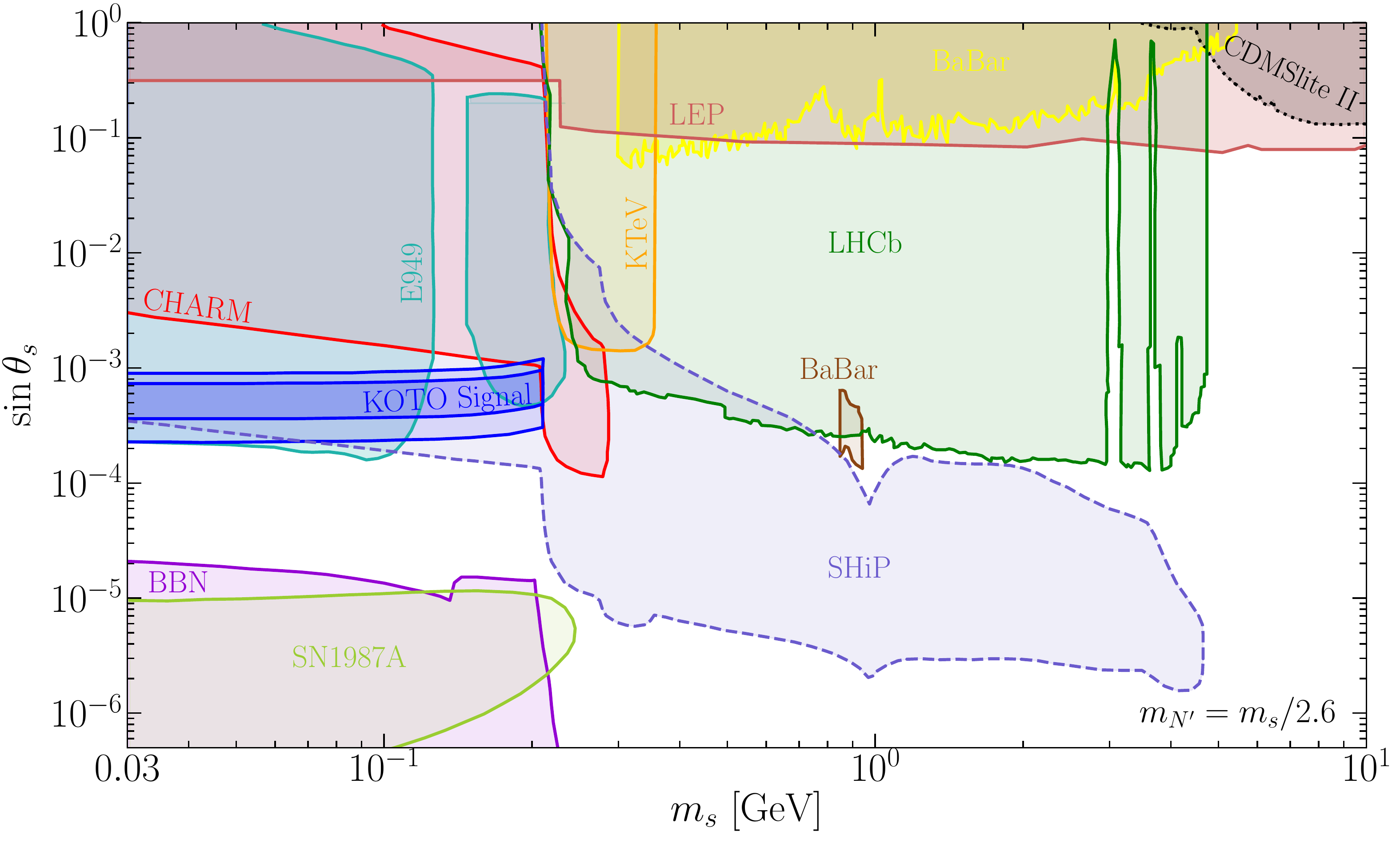} 
 \caption{Constraints on a light singlet mediator, in the
$m_s$-$\theta_s$ plane, for the case $m_s = 2.6\, m_{N'}$. 
The experimental bounds, along with the projected sensitivity, are the same as in Figure\ \ref{fig:koto} and taken from
 ref.\ \cite{Winkler:2018qyg}. The strongest direct detection constraint derived for our model comes from CDMSlite II experiment~\cite{Agnese:2015nto} and is shown with the black dotted line.
 } 
 \label{fig:kotoOther}
\end{center} 
\end{figure*}

The mixing of $s$ with the Higgs boson leads to 
spin-independent DM interactions
with nucleons. In particular, the cross section for scattering
on nucleons is
\begin{equation}
\sigma_\mathrm{SI}^p=\frac{g_s^2m_{N^\prime}^2m_n^4\,\sin^22\theta_s\,
f_n^2}{4\pi\left(m_{N^\prime}+m_n\right)^2\bar
v^2}
  \left(\frac{1}{m_h^2}-\frac{1}{m_s^2}\right)^2,
\label{ddeq}
\end{equation}
where $m_n=0.938~\mathrm{GeV}$ is the nucleon mass, 
$f_n=0.30$ is the relative coupling of the Higgs to nucleons
\cite{Cline:2013gha,Ellis:2018dmb}, $m_h=125~\mathrm{GeV}$
is the SM-like Higgs mass, $m_s$ is the singlet mass
and $\theta_s$ is the $s$-$h$ mixing angle.  (Recall that $\bar v\cong
174\,$GeV is the complex Higgs field VEV.)

The strongest constraints from direct detection, in the mass
range $m_{N'}< 4.5$\,GeV predicted in our model, come from the
experiments CRESST II~\cite{Angloher:2015ewa}, 
CDMSlite II~\cite{Agnese:2015nto} and LUX~\cite{Akerib:2015rjg}.
In the future these limits will be improved by 
SuperCDMS \cite{Agnese:2016cpb}. In all cases the sensitivity 
rapidly drops with lower DM mass because
of the threshold for energy deposition.  The DarkSide experiment
\cite{Agnes:2018ves} claims limits below those mentioned above,
but their validity has been questioned in ref.\ 
\cite{Bernreuther:2019pfb}, and we omit them from our analysis.

Recently ref.\ \cite{Wu:2019nhd} cast doubts on the robustness
of direct constraints on light dark matter in light of astrophysical
uncertainties, especially that of the local escape velocity, that has been
revisited using Gaia data \cite{2019MNRAS.485.3514D}.  It is claimed that the 2017 cross
section bound from XENON1T \cite{Aprile:2017iyp} at DM mass $4$\,\,GeV is uncertain by six
orders of magnitude.  We checked their results using DDCalc 
\cite{Workgroup:2017lvb},
finding only two orders of magnitude uncertainty.  More importantly,
the astrophysical uncertainty on the more relevant newer constraints
is only a few percent (due to the much lower thresholds of those
experiments), hence irrelevant.

For a given value of DM mass $m_{N'}$, we can use the relic density
constraint shown in Figure\ \ref{fig:conts} to 
determine the coupling $g_s$.  Then the predicted direct detection
cross section (\ref{ddeq}) leads to a constraint in the
$m_s$-$\theta_s$ plane, that we plot as a dashed curve in
Figure\ \ref{fig:koto} for the case $m_s < m_{N'}$.  As mentioned above, for larger values of $m_{N'}$ the direct detection
constraint is stronger, and the region favored by the KOTO anomaly is excluded.

For $m_s > m_{N'}$, and particularly in the region where $m_s \gtrsim 2\, m_{N'}$, the direct detection constraints are shown in Figure\ \ref{fig:kotoOther} for the case considered in the center panel of Figure\ \ref{fig:conts}, namely for $m_s = 2.6\,m_{N'}$. These bounds are much weaker than those in Figure\ \ref{fig:koto} for any value of the DM mass $m_{N'}$ because of the relatively larger assumed value of $m_s$, as can be seen from eq.\ (\ref{ddeq}).

\subsection{DM indirect detection}
Light dark matter models are typically constrained by indirect
signals, like annihilation in the galaxy or the cosmic microwave
background (CMB), enhanced by the relative large abundance of 
light DM.  These signals are suppressed for asymmetric dark matter,
by the absence of the symmetric component with which to annihilate,
but DM accumulation in stars can provide significant constraints in
the asymmetric case.  Our model provides for a continuum of
possibilities between the purely symmetric and asymmetric cases,
depending on the strength of the coupling $g_s$ when $N'\bar N'\to s
s$ is the dominant process, or a combination of $g_s$ and $m_s$ when
$s$-channel annihilation dominates (recall Figure\ \ref{fig:conts}).

However in our scenario there are several reasons for annihilation
signals to be suppressed at late times, even in the symmetric regime.
For the case where $N'\bar N'\to s s$ dominates, the cross section is $p$-wave, which significantly relaxes indirect constraints because of the low DM velocity at times much later than freezeout \cite{Diamanti:2013bia}.  An exception can occur if the DM particles annihilate to form a bound state \cite{An:2016kie}, which is $s$-wave, and leads to much stronger CMB constraints than the $p$-wave process.  However this only occurs for relatively heavy DM, with mass $\gtrsim 250\,$GeV.  

In addition, the $p$-wave process we consider from $N'\bar
N'\to f\bar f$ targets parameter space with $m_s > 2\, m_{N'}$, such
that $m_s$ is not too close to the threshold $2\, m_{N'}$.
In this case the indirect signal is further suppressed, due to the low DM velocity in galaxies, $v \sim 10^{-3}\,\,c$, since the phase-space average of the annihilation cross section samples the resonant region much less than in the early universe during freezeout. 
Indeed, following refs.\ \citep{Diamanti:2013bia,Cline:2019okt}, it is possible to estimate that for the values of $g_s$ and $m_{N'}$ contained in Figure\ \ref{fig:conts} (center and right plots) the maximum ratio between the DM annihilation cross section today and that at the time of freezeout, given by eq.\ \eqref{sigmavNWA}, is of order of $\sim 10^{-14}$, which leads today to $\langle \sigma v \rangle \lesssim 10^{-37}\,\,\text{cm}^3/\text{s}$. Such a value is well below the most stringent indirect detection constraint for $p$-wave annihilating DM of mass $m_{N'} \lesssim 4.5$ GeV \cite{Diamanti:2013bia}.

Another possible signal that does not rely upon DM annihilation with itself, but rather on its interactions with standard model particles, is the effect of DM accumulation within stars.
The most promising sites for capturing  DM  are neutron stars (NS's) because of their high density, which enhances the probability for DM particles to be captured and accumulate in the center of the NS during its lifetime. 

For purely asymmetric DM, there is no DM annihilation in the NS core, and its accumulation may start to increase the star mass, destabilizing the delicate balance between the gravitational attraction and the Fermi pressure, and leading to the gravitational collapse of the NS into a black hole \citep{Bertone:2007ae,Guver:2012ba}. However, this effect is only relevant in the case of bosonic DM, where there is
no compensating increase in the Fermi pressure, leading to
 severe constraints on the DM-nucleon and DM-lepton scattering cross-sections based on the estimated age of the oldest NSs observed so far \citep{Bertoni:2013bsa,Alonso-Alvarez:2019fym}.
These bounds do not apply to the present model because of the fermionic nature of our DM candidate, and its GeV mass scale. For fermionic asymmetric DM, the destabilization  can occur only for DM with mass larger than the PeV scale
and having attractive self-interactions
\cite{Gresham:2018rqo}.

In the case where DM is partially or purely symmetric, which occurs for smaller values of $m_{N'}$ and $g_s$ in our model (recall Figure \ref{fig:conts}), the accumulated DM inside the NS core can annihilate and the annihilation products might thermalize, heating up the star and contributing to its luminosity \cite{Baryakhtar:2017dbj}.
The latter is also increased by DM kinetic heating from multiple DM scatterings with the NS constituents, namely neutrons, electrons and muons, and this effect is independent of whether the DM is symmetric or asymmetric \citep{Baryakhtar:2017dbj,Bell:2019pyc}. However, the expected NS surface temperature generated only by DM annihilation and scattering is too low to be detected by current infrared telescopes. A future detection by, for instance, the James Webb Space Telescope, would set the strongest bound on the DM-nucleon and DM-lepton scattering cross-sections for DM masses below the GeV scale, which would constrain our model \cite{Bell:2019pyc}.

Other limits on DM-nucleon interaction can in principle be derived from DM capture by white dwarfs (WD's)\ \cite{Bertone:2007ae}. Similarly to NS's, asymmetric DM accumulating in the WD core might destabilize the latter and spark fusion reactions that precede a Type Ia supernova explosion\ \cite{Bramante:2015cua}. However, in models where DM interactions with SM particles occur only via a light scalar mediator mixing with the Higgs boson, destabilization effects become important only for fermionic DM masses above $10^6$ GeV\ \cite{Acevedo:2019gre}.

On the other hand, DM scattering and annihilation can heat up WD's and contribute to their luminosity. The difference between the WD and the NS case is that very old WD's with low surface temperature have been observed, in particular within the M4 globular cluster \citep{Richer:1995wg,Bedin:2009}. Such observations have been used to claim very strong constraints on the DM-nucleon scattering cross section,  $\sigma_{SI} \lesssim 10^{-42} - 10^{-43}\,\,\text{cm}^2$ for DM masses in the range $10^{-2} - 10^{7}\,\,\text{GeV}$ \citep{McCullough:2010ai,Dasgupta:2019juq}. These limits were derived based on the assumption that the DM density within the M4 globular cluster is as high as $10^3\,\,\text{GeV}/\text{cm}^3$, which can make the DM contribution to the WD luminosities as high as the observed
values. However, as pointed out by ref.\ \cite{McCullough:2010ai}, the value of the DM density in globular clusters is highly uncertain and under debate. Although values of $10^3\,\,\text{GeV}/\text{cm}^3$ could be expected if globular clusters form within DM subhalos before falling into galactic halos \cite{Peebles:1984}, tidal stripping by subsequent mergers\ \cite{Griffen:2010}  provides a very efficient way of depleting DM in these systems, leaving them dominated today by just the stellar component \cite{Beasley:2020}.
The observation that the present-day dynamics of globular clusters can be explained without the need of DM suggests that these systems might form in molecular clouds in the gaseous disk of the galaxy instead of in DM overdensities \citep{Kravtsov:2003sm,Hooper:2010es,Claydon:2019}. 
It is therefore reasonable to assume that the DM density in the M4 globular cluster, which is about 2 kpc away from us in the direction towards the galactic center, could be as low as in the solar neighborhood, $\sim 0.3\,\,\text{GeV}/\text{cm}^3$. This lower value would reduce the saturated DM heating luminosity by approximately three orders of magnitude, well below the observed one, and lead to no bound on the DM-nucleon scattering cross section at all. 
More promising WD candidates might be found in globular clusters in dwarf spheroidal galaxies of the Milky Way, where a significant amount of DM may have survived tidal stripping \cite{Krall:2017xij}.

\begin{figure}[t]
\begin{center}
\centerline{\includegraphics[scale=0.315]{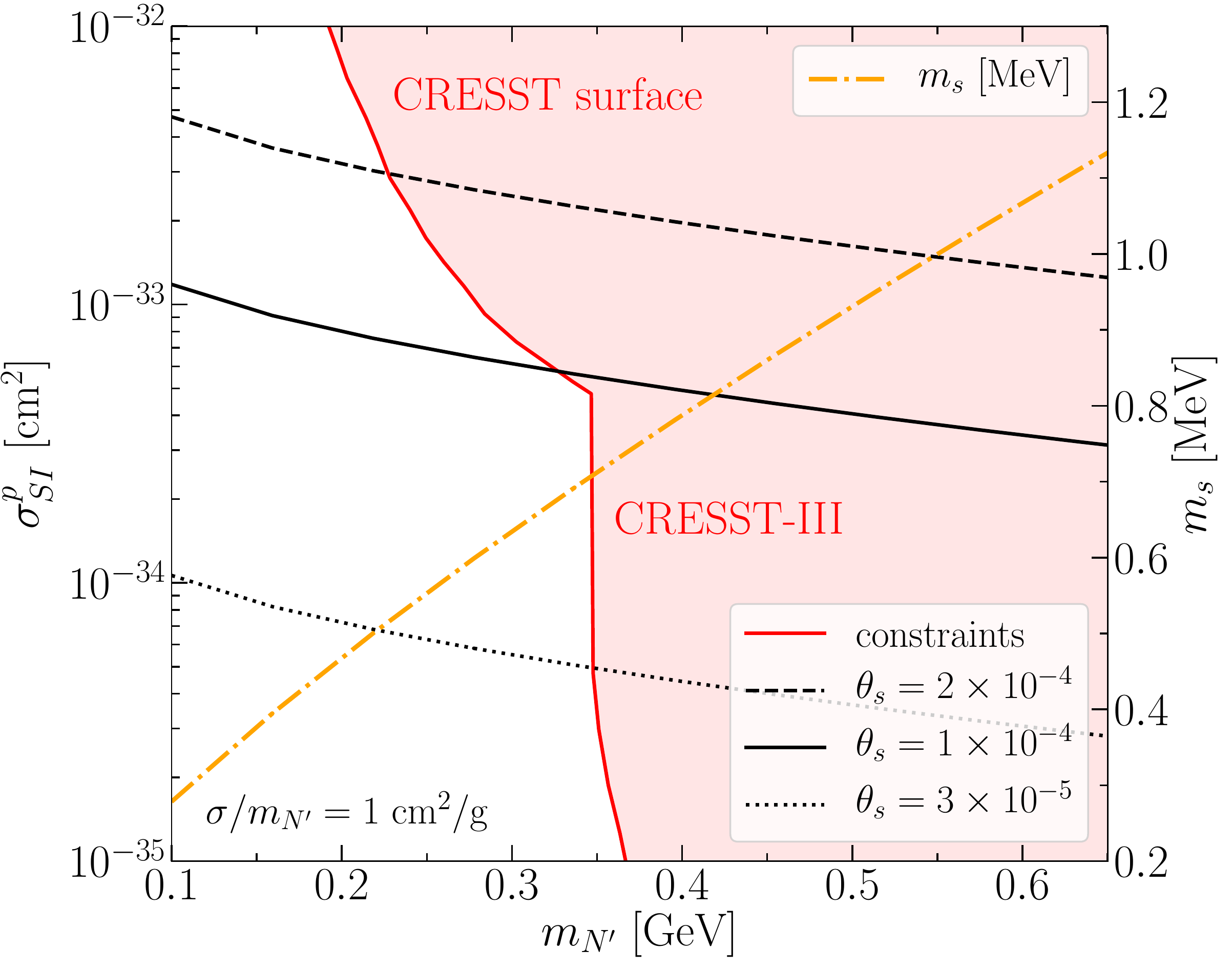}
\hfil \includegraphics[scale=0.31]{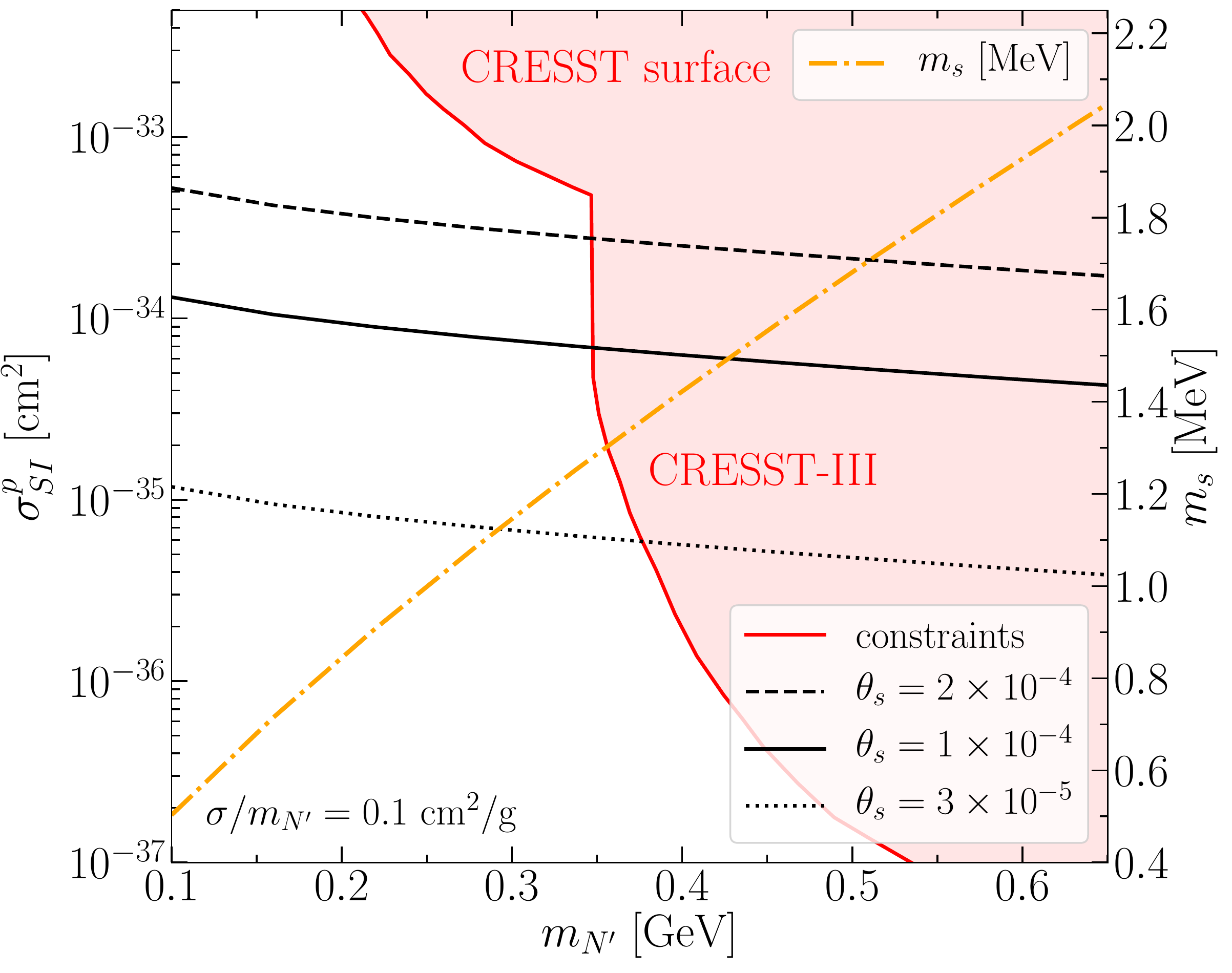}}
 \caption{Predicted spin-independent cross section for DM scattering
on nucleons versus DM mass $m_{N'}$, assuming approximately symmetric
DM with a self-interaction cross section of
$\sigma/m_{N'}$ = 1\,cm$^2$/g (left) or 0.1\,cm$^2$/g (right), for three choices of $\theta_s$
(dashed, solid black, dotted) and the envelope of experimental
constraints (with the exception of DarkSide-50) copied from ref.\ 
\cite{Akerib:2018hck} (solid red).  Dash-dotted curve shows the
singlet mass $m_s$ versus $m_{N'}$.} 
 \label{fig:dd}
\end{center} 
\end{figure}

\subsection{DM self-interactions}

Dark matter can also interact with itself by exchange of the $s$,
which is of interest for addressing small-scale structure problems of
collisionless cold dark matter (see ref.\ \cite{Tulin:2017ara} for a
review).  Ref.\ \cite{Tulin:2012wi} showed that the self-interaction
cross section can be at an interesting level for solving these
problems, while obtaining the right DM relic density, if both 
$m_{N'}$ and  $m_s$ are light,
\be
   m_{s} \cong 1\,\,{\rm MeV}\times \left\{\begin{array}{ll}
\left(m_{N'}\over 0.55\,\,{\rm
GeV}\right)^{3/4},  & \sigma/m_{N'} = 1\,\,{\rm cm}^2/{\rm g}\\
\left(m_{N'}\over 0.25\,\,{\rm
GeV}\right)^{3/4},  & \sigma/m_{N'} = 0.1\,\,{\rm cm}^2/{\rm g}
	\end{array}\right.
\label{mNms}
\ee
These relations, valid for approximately symmetric DM, correspond to
self-interaction cross section per mass in the range
$\sigma/m_{N'} = 0.1-1\,$cm$^2$/g, that are relevant for suppressing
cusps in density profiles of dwarf spheroidal to Milky Way-sized
galaxies.

Such light singlets in the MeV mass range are strongly constrained 
by direct detection.  The prediction (\ref{ddeq})
is modified by the fact that the momentum transfer $q$ is no longer
negligible compared to $m_s$, hence $m_s^2\to m_s^2 + q^2$ in 
eq.\ (\ref{ddeq}).  We take $q = m_{N'}v_{N'}$ with DM velocity
$v_{N'}\sim 300\,\,$km/s to account for this.  
Figure\ \ref{fig:koto} shows that for low $m_s$
there is an allowed window for $\sin\theta_s\sim 2\times (10^{-5}-
10^{-4})$ between the BBN and E949 constraints, which persists to 
smaller values of $m_s\gtrsim 1$\,\,MeV before being  excluded by
BBN as $m_s$ falls below the threshold for $s\to e^+e^-$ decay.

In Figure\ \ref{fig:dd} we show the predicted spin-independent cross
section versus $m_{N'}$ (black curves) for several choices of $\theta_s$ in the experimentally 
allowed range, fixing $g_s$ as in Figure\ \ref{fig:conts} to give the right
relic density, and $m_s$ (orange dash-dotted curve) as a function of $m_{N'}$ using
(\ref{mNms}).   
The plot on the left assumes the higher self-interaction cross section $\sigma/m = 1\,$cm$^2$/g.  In this case, it is necessary to take the singlet mass $m_s\lesssim 0.7\,\,$MeV 
and the DM mass $m_{N'}\lesssim 0.3\,\,$GeV to respect the
direct detection limit.  This is ruled out by BBN since the
decay $s\to e^+e^-$ are kinematically blocked, and will lead to overclosure by the singlets as $T$ falls below $m_s$.
However, by adopting a lower self-interaction cross section
$\sigma/m = 0.1\,\,$cm$^2$/g (right plot of Figure\ \ref{fig:dd}), which may still be relevant for
some of the small-scale structure issues, the allowed range of
$m_{N'}$ and $m_{s}$ is increased to somewhat larger values
with $m_s > 2\, m_e$, which can be compatible with BBN.

The previous determination holds in the region $m_{N'}\lesssim 3\,$GeV where the DM is to a good approximation symmetric, corresponding to the linearly increasing
branch of the relic density contour in Figure\ \ref{fig:conts} (left).  For
nearly asymmetric DM, the horizontal branch with $m_{N'} \cong
4.5\,\,$GeV applies.  Instead of eq.\ (\ref{mNms}), the desired
self-interaction cross section requires a roughly
 linear relation $g_s\cong
0.75 + 4.43\, m_s/$GeV (valid for $m_s\sim 0.2-0.3\,\,$GeV), 
that we determine by applying a Sommerfeld
enhancement factor \cite{Cassel:2009wt} to the tree-level, phase-space averaged transport scattering cross
section given in ref.\ \cite{Tulin:2012wi}, and requiring that the resulting cross
section is $\sigma/m_{N'} = 1\,\,$b/GeV $\cong 0.6\,\,$cm$^2$/g for a mean DM velocity of $10$\,
km/s, corresponding to dwarf spheroidal galaxies.  To satisfy the
\nobreak{CDMSLite II} constraint $\sigma_{SI}<1\times 10^{-41}$\,cm$^2$ at
$m_{N'}= 4.5\,\,$GeV \cite{Jiang:2018pic}, it is necessary to take
small mixing $\theta_s\lesssim 6\times 10^{-6}$, since $g_s\sim 2$
for $m_s\sim 0.2-0.3\,\,$GeV, from imposing the desired
value of $\sigma/m_{N'}$.  

Hence we find two allowed regions for strong self-interactions,
one marginal since $m_s\gtrsim 1\,\,$MeV close to BBN limits, with
$\sigma/m_{N'}\sim 0.1\,\,$cm$^2$/g at the low end of the range desired
for small scale structure, and $m_{N'}\sim 0.35\,\,$GeV.  
The other allows for a larger $\sigma/m_{N'}\gtrsim 0.6\,\,$cm$^2$/g,
with singlet parameters close to the SN1987A exclusion curve and
$m_{N'}\cong 4.5$\,\,GeV.

\section{Naturalness}
\label{naturalness}

In our proposal, the flavor structure of neutrinos is controlled by
the same matrix $\eta_{\nu, ij}$ that governs the HNL
couplings, up to a proportionality constant, in the spirit of MFV.
In order for DM to be stable, $\eta_{\nu, ij}$ must have rank two.
The HNL mass matrix is proportional to the identity, up to corrections
going as $\eta_\nu^2$.  
We do not provide any fundamental explanation of the origin of this
structure; instead we content ourselves with the feature that it is
technically natural in the sense of 't Hooft: all radiative
corrections are consistent with our assumptions.

The stability of DM is most easily seen in the basis 
(\ref{NHeta}, \ref{IHeta}), where $N'$ obviously decouples from the
SM leptons.  We assume this coincides with the mass
eigenbasis, which is consistent since there are no interactions that
can induce mass-mixing between $N'$ and the remaining $N_i$'s.
Self-energy corrections involving $s$ exchange are flavor-diagonal.  
Those involving Higgs and leptons in the loop leave $m_{N'}$ unchanged,
while renormalizing the $N_i$ mass matrix by
\be
	M_N\delta_{ij} \to M_N\delta_{ij} + O(1)\times \eta_{\nu,ik}\,
	{m_{\ell_k}\over 16\pi^2}\, \eta^{\dag}_{\nu,kj}
\label{Msplit}
\ee
where $m_{\ell_i}$ are the charged lepton masses.  Given the smallness
of $\eta_\nu \lesssim 10^{-4}$, these corrections are unimportant.
Similarly the one-loop corrections to $\eta_{\nu}$ are negligible,
\be
	\eta_{\nu} \to \eta_{\nu} + {O(1)\over 16\pi^2}\,\eta_{\nu}
	\eta_{\nu}^{\dag} \eta_{\nu}
\ee
and cannot induce couplings to $N'$.  The only particles to which
$N'$ couples are the singlet and the inflaton, eqs.\ (\ref{Nphi}, \ref{eq:s_int}), and these 
interactions are assumed to be flavor-conserving at tree level.
Flavor-changing corrections to $g_s$ and $g_\phi$ of
$O(\eta_\nu^2/(16\pi^2))\times g_{s,\phi}$ arise at the one-loop level and
are negligible for our purposes.
 
There remains the infamous naturalness problem of the Higgs
mass (weak scale hierarchy).  This problem was addressed in the
context of the seesaw mechanism in ref.\ \cite{Brivio:2017dfq}, where
the weak scale was linked to that of the heavy Majorana neutrinos
by radiative generation of the Higgs potential.  A low scale for
their masses is needed, $M_{\nu_R} \lesssim 10^7$\,\,GeV 
\cite{Brivio:2018rzm}, which would
require a low reheat temperature in our scenario, and
consequently small coupling $g_\phi \lesssim 10^{-8}$.  Although
peculiarly small, this value would still be compatible with the
requirements of technical naturalness since it can only be
multiplicatively renormalized.  

The very light singlet could pose an analogous problem of fine-tuning.
The first threshold encountered when running the renormalization
scale up from low values is that of $N_i$, which contributes of order
$\delta m_s \sim g_s M_N/(4\pi)$ to $m_s$.  This can easily be
compatible with the tree-level values of $m_s$ desired for large
parts of the allowed parameter space (see Figures\ \ref{fig:conts} 
and \ref{fig:koto}).  

Next one encounters the Higgs
threshold, which further shifts $m_s$ through 
the coupling $\lambda_{hs}$.  The correction is of order $\delta m_s \sim {\sqrt{\lambda_{hs}}\, m_h/4\pi}$
which is related to the mixing angle by $\theta_s\sim \lambda_{hs} v
v_s/m_h^2$, where $v$ and $v_s$ are the respective VEVs of the Higgs
and the singlet.  In turn, $v_s$ depends upon the $s$ self-coupling
through $m_s^2 \sim \lambda_s v_s^2$.  Using these and demanding that
$\delta m_s\lesssim m_s$ gives the constraint 
$\sqrt{\lambda_s} \lesssim  {16\pi^2m_s^3v/(\theta_s m_h^4)}$.
This can always be satisfied by choosing small enough $\lambda_s$,
but the latter has a minimum natural value given by its 
one-loop correction $\delta\lambda_s \sim
g_s^4/(16\pi^2)$.\footnote{There is also a one-loop correction of 
order $\lambda_{hs}^2/16\pi^2$, but this leads to a weaker bound
on $\theta_s$ than (\ref{thbound}).}\ \ 
 Putting all of these together, we get a
naturalness bound on the singlet mixing angle
\be
	\theta_s\lesssim \left(4\pi\, m_s\over m_h\right)^{3}
	\left(1\over \sqrt{\lambda_h}\,g_s^{2}\right)\sim 0.008
\label{thbound}
\ee
(taking $m_s\sim 0.3\,\,$GeV and $g_s\sim0.1$)
which is compatible with the regions of interest for future
discovery, including the anomalous KOTO events.
Thus, somewhat surprisingly, the light scalar does not introduce
a new hierarchy problem analogous to that of the Higgs mass,
due to its relatively weak couplings.

We do not address the smallness of $\theta_{QCD}$ in our ``theory
of everything,'' which was a motivation for refs.\ 
\cite{Salvio:2015cja,Ballesteros:2016euj} to choose the QCD axion as their dark
matter candidate.  This neglect is consistent with our philosophy of
focusing on technical naturalness rather than aesthetic values of
couplings, since $\theta_{QCD}$ is known to be highly stable against
radiative corrections \cite{Ellis:1978hq}.

\section{Conclusions}
\label{conclusions}

It is interesting to construct scenarios that link the
different particle physics ingredients known to be missing from the
standard model, since it can lead to distinctive predictions.
Here we have constructed a minimal scenario that explains inflation,
baryogenesis, dark matter and neutrino masses, is highly
predictive, and can be tested in numerous experimental searches for
heavy neutral leptons, light dark matter, and light scalar mediators.
At low energies, the only new particles are three quasi-Dirac HNLs,
one of which is DM (and exactly Dirac), and a light singlet scalar.

One prediction of the model is that no new source of CP-violation  is
required for baryogenesis, which occurs through a novel form of
leptogenesis here.  In contrast to ordinary leptogenesis, the
asymmetry is formed during inflation, and the right-handed
neutrinos that generate light neutrino masses are too heavy to be
produced during
reheating.  CP is spontaneously broken by the inflaton VEV
during inflation, and the light HNLs transmit the lepton asymmetry
from the inflaton to the SM.  In ref.\ \cite{Cline:2019fxx} it was shown that
observable isocurvature perturbations can arise, depending on the
inflaton potential and initial conditions.  In the present model,
these would appear as correlated dark matter isocurvature and
adiabatic perturbations. 

Another prediction is that the two unstable HNLs $N_i$ should be degenerate to very 
high precision with the dark matter $N'$, split only by the correction
(\ref{Msplit}) of order $10^{-2}$\,\,eV.  Similarly, the $N_i$ are Dirac
particles to a very good approximation, with a lepton-violating
Majorana mass of order $10^{-6}$\,\,eV.  This is too small to be 
detectable in neutrinoless double beta decay, but large enough to
allow for a distinctive signature of lepton
violation through $N$-$\bar N$ oscillations.
The two $N_i$ HNLs can mix strongly enough with SM
neutrinos to be discoverable at upcoming experiments like SHiP. 
The stability of $N'$ is directly linked to
the masslessness of the lightest neutrino.
This
connection could be relaxed by slightly modifying the assumption
that the HNL couplings are aligned with light neutrino masses through
eq.\ (\ref{epsrel}), without spoiling other features of our model.   We further showed that lepton-flavor-violating decays like $\mu\to e\gamma$ and $\mu\to 3e$ may be generated by HNL exchange in loops, at a level that can be detected
in future experiments.

In our framework, the dark matter $N'$ is partially asymmetric, and has a
mass bounded by $m_{N'} \lesssim 4.5\,\,$GeV.  The bound is saturated
when $N'$ is purely asymmetric, and its mass is determined by the 
observed value of the baryon asymmetry.  Light DM can be accommodated
by taking small values of the coupling $g_s$ between $N'$ and the 
singlet $s$, which controls $N'\bar N'\to s s$ annihilation;
see Figure\ \ref{fig:conts}.  In the
mass range $(1-4.5)\,\,$GeV, significant constraints are already placed
by direct DM searches.  

The light scalar singlet, whose mass must be less than $m_{N'}$ for
efficient $N'\bar N'\to s s$ annihilation,  can lead to striking 
signatures.  For example the decay $K_L\to \pi s$ can explain
anomalous excess events recently observed by the KOTO experiment,
but only if $m_{N'}\lesssim 2.5\,\,$GeV; otherwise direct detection
constraints rule out this mode at the level suggested by the KOTO
events, where $m_s\sim (100-200)$\,\,MeV and $s$ mixes with the Higgs
at the level $\theta_s\sim 5\times 10^{-4}$. (The preferred
parameter region for the KOTO anomaly is only a small part of the
full allowed space of our model.)
In a different part of parameter space with $m_s \sim (0.2-0.3)$\,\,GeV,
$m_{N'}\cong 4.5\,\,$GeV 
and $\theta_s\lesssim 6\times 10^{-6}$, the singlet mediates DM
self-interactions with a cosmologically interesting cross section,
$\sigma/m_{N'}\sim 0.6$\,\,cm$^2$/g.

\acknowledgments
We thank J.\ Kopp for an insightful
question that inspired this work, and G.\ Alonso-\'{A}lvarez, J.\ Bramante, T.\ Bringmann, L.\ Di Ciaccio, M. Fairbairn,
E.\ Fernandez-Martinez, F.\ Kahlhoefer, C.\ Lacasta, 
E.\ Migliore,   D.\ Morrissey,  S.\ Petcov, M. Reina-Campos, P.\ Scott, J.-L.\ Tastet and J.\ Timmermans 
   for helpful correspondence.
Our research is supported by NSERC (Natural Sciences and Engineering
Research Council, Canada).  MP is supported by the Arthur B.\ McDonald
Institute for Canadian astroparticle physics research.\\

\appendix
\section{Decay rate for $N_{i} \to \nu \ell^{+} \ell^{-}$}
\label{appA}

The matrix element for the process $N_{i} \to \nu_{\beta} \ell^{+}_{\beta} \ell^{-}_{\alpha}$, where $\alpha, \beta = e$, $\mu$, $\tau$, is
\be
\mathcal{M} = \frac{g_w^2}{8 M_W^2}\,\Big[\bar{u} (p_{\ell_{\alpha}^{-}})\, \gamma^{\mu} (1 -\gamma^5)\,u (p_{N_i})\,U^{\ast}_{i\alpha} \Big]\,
\Big[\bar{u} (p_{\nu_{\beta}})\, \gamma_{\mu} (1 -\gamma^5)\,u (p_{\ell_{\beta}^{+}}) \Big]
\ee
whose square reduces to
\be
\langle |\mathcal{M}|^2 \rangle = \frac{G_F^2}{16} |U_{i\alpha}|^2 M_{i} E_{\beta} \bigg[\frac{M_i^2 + m_{\beta}^2 - m_{\alpha}^2}{2} - M_i E_{\beta} \bigg]
\label{M2}
\ee
after averaging over the initial spin, summing over final spins and setting $m_{\nu_{\beta}} = 0$. Here, $G_{F}$ is the Fermi constant, $E_{\beta}$ is the energy of $\ell_{\beta}^{+}$ and we have defined for simplicity $M_{i} \equiv M_{N_i}$, $m_{\alpha} \equiv m_{\ell_{\alpha}^{-}}$ and $m_{\beta} \equiv m_{\ell_{\beta}^{+}}$. 
The decay rate $\Gamma$ can be obtained by plugging eq.\ \eqref{M2} in the standard decay formula (see ref.\ \cite{Tanabashi:2018oca}) and computing the three-body phase space integral.
The common assumption made in the literature is to consider $m_{\beta} = 0$, which is well motivated for $\alpha = e,\,\,\mu$ and $\beta = \mu,\,\,e$. In these cases, the decay rate is\ \cite{Johnson:1997cj, Gorbunov:2007ak}
\be
\Gamma = \frac{G_F^2 M_{i}^5}{192 \pi^3}\,\, |U_{i\alpha}|^2\,\,
\Big(1 - 8\, x_{\alpha}^2 + 8 x_{\alpha}^6 - x_{\alpha}^8 - 12\, x_{\alpha}^4\, \log (x_{\alpha}^2) \Big)
\label{Gammam0}
\ee
where $x_{\alpha} = m_{\alpha} / M_{i}$.
Such a simplified formula does not hold for $\alpha = \mu,\,\,\tau$ and $\beta = \tau,\,\,\mu$, where the muon mass is not negligible compared to the tau mass.
The general expression reads
\be
\begin{split}
\Gamma &=\frac{G_F^2 M_i^5}{192 \pi^3}\,\,|U_{i\alpha}|^2\,\,\Bigg\{12\, |x_{\beta}^2-x_{\alpha}^2|\, (x_{\beta}^2+x_{\alpha}^2)\, \\
&\quad \log\Bigg[\frac{x_{\beta}^2+x_{\alpha}^2 - (x_{\beta}^2-x_{\alpha}^2)^2 - |x_{\beta}^2-x_{\alpha}^2|\,\sqrt{(1-(x_{\beta}-x_{\alpha})^2) (1-(x_{\beta}+x_{\alpha})^2)}}{2\, x_{\beta}\, x_{\alpha}}\Bigg] \\
&-12\, \Big[x_{\beta}^4+x_{\alpha}^4-2 x_{\beta}^4\, x_{\alpha}^4\Big] \log\Bigg[\frac{1 - x_{\beta}^2 - x_{\alpha}^2 - \sqrt{(1-(x_{\beta}-x_{\alpha})^2) (1-(x_{\beta}+x_{\alpha})^2)}}{2\, x_{\beta}\, x_{\alpha}}\Bigg] \\
&+\sqrt{(1-(x_{\beta}-x_{\alpha})^2)(1-(x_{\beta}+x_{\alpha})^2)} \\
&\quad \bigg[1 - 7\, \Big(x_{\beta}^2 + x_{\alpha}^2\Big) \Big(1 + x_{\beta}^2\,x_{\alpha}^2\Big) -7\, \Big(x_{\beta}^4 + x_{\alpha}^4\Big) + 12\, x_{\beta}^2\, x_{\alpha}^2 + x_{\beta}^6 + x_{\alpha}^6\bigg]\Bigg\}
\end{split}
\ee
where $x_{\alpha} \equiv m_{\alpha} / M_{i}$ and $x_{\beta} \equiv m_{\beta} / M_{i}$. It is easy to check that this formula reduces to eq.\ \eqref{Gammam0} in the limit $m_{\beta} \to 0$.

\end{document}